\documentclass[a4paper,12pt]{article}

\usepackage[utf8]{inputenc}
\usepackage[T1]{fontenc}

\usepackage[english]{babel}

\usepackage{newtxtext,newtxmath}

\usepackage{scicite}

\usepackage{graphicx}
\usepackage[labelsep=period,labelfont=bf,small]{caption}

\usepackage[left=2.5cm,right=2.5cm,top=2.5cm,bottom=2.5cm]{geometry}

\usepackage{titling}
\usepackage{sectsty}
\sectionfont{\normalsize\bfseries}
\subsectionfont{\normalsize\rm\itshape}

\usepackage{authblk}

\usepackage{lineno}

\usepackage[hidelinks]{hyperref}
\usepackage{setspace}
\usepackage{framed,color}
\definecolor{myorange}{RGB}{253, 184, 99}
\definecolor{mygreen}{RGB}{13, 100, 13}
\definecolor{mypurple}{RGB}{140, 0, 211}

\newcommand{\newtext}[1]{#1}

\date{\normalsize \today}
\title{\newtext{Geospatial distributions reflect temperatures of linguistic features}}

\author[1]{Henri Kauhanen\thanks{Corresponding author: henri.kauhanen@uni-konstanz.de}}
\author[2]{Deepthi Gopal} 
\author[3,4]{Tobias Galla}
\author[5]{Ricardo Berm\'udez-Otero}
\affil[1]{Zukunftskolleg, University of Konstanz, Universit\"atsstra{\ss}e 10, 78464 Konstanz, Germany}
\affil[2]{Department of Theoretical and Applied Linguistics, University of Cambridge, Sidgwick Avenue, Cambridge, CB3 9DA, UK}
\affil[3]{Department of Physics and Astronomy, School of Natural Sciences, The University of Manchester, Oxford Road, Manchester, M13 9PL, UK}
\affil[4]{Instituto de F\'isica  Interdisciplinar y Sistemas  Complejos, IFISC (CSIC-UIB), Campus Universitat Illes Balears,  E-07122 Palma de Mallorca, Spain}
\affil[5]{Department of Linguistics and English Language, The University of Manchester, Oxford Road, Manchester, M13 9PL, UK}
\begin{document}

\maketitle

\section*{Short title}

\newtext{Temperatures of linguistic features}

\section*{One-sentence summary}

\newtext{The propensity of features of human language to undergo cultural change can be measured by a temperature-like quantity, whose value can be estimated from those features' present-day geospatial distributions.}

\section*{Abstract}

\newtext{Quantifying the speed of linguistic change is challenging due to the fact that the historical evolution of languages is sparsely documented. Consequently, traditional methods rely on phylogenetic reconstruction. In this paper, we propose a model-based approach to the problem through the analysis of language change as a stochastic process combining vertical descent, spatial interactions, and mutations in both dimensions. A notion of linguistic temperature emerges naturally from this analysis as a dimensionless measure of the propensity of a linguistic feature to undergo change. We demonstrate how temperatures of linguistic features can be inferred from their present-day geospatial distributions, without recourse to information about their phylogenies. Thus the evolutionary dynamics of language, operating across thousands of years, leaves a measurable geospatial signature. This signature licenses inferences about the historical evolution of languages even in the absence of longitudinal data.}

\section*{Keywords}

cultural evolution; \newtext{languages; linguistic stability; linguistic temperature}

\section*{Introduction}

Since the biological emergence of modern language some 100,000 years ago \cite{Bic2007}, human languages have diversified through processes of cultural evolution to the extent that thousands of distinct languages are spoken today around the world \cite{Pag2009}. These languages display an enormous amount of variation in a combinatorial space spanned by a finite number of structural features, whose possible values emerge from biological and cognitive constraints on linguistic representation and language use. These features determine how individual words are formed, how words are combined into phrases and sentences, and which sounds and sound sequences are available in any given language.

\newtext{The causes of linguistic change have been debated ever since the birth of modern linguistic theory in the late 19th century, and a number of these processes are now understood in detail \cite{Lab2007}. The most basic general insight emerging from this work, translated into terms that are current in the study of evolution in other fields \cite{JainEtAl1999}, is that language change is both vertical and horizontal. Under ordinary circumstances, language is relatively reliably passed on from parents to children, which accounts for the vertical, intergenerational descent of linguistic features across phylogenetic lineages. It is possible for this transmission to fail, however, and for a feature to change, a process not unlike point mutations in the genome---although debate exists over whether linguistic mutations are mostly random or directed \cite{NewEtAl2017}. It is also possible for the vertical line of descent to be interrupted by horizontal effects, in which a feature of a language changes due to the influence of a phylogenetically-distinct but geographically-neighbouring language; the empirical problem of distinguishing the results of horizontal effects from the results of failed vertical transmission recurs often in many areas of historical linguistics \cite{Lab2007}.}

\newtext{Within the study of the dynamics of language, there is a large and rich body of work that seeks to measure the susceptibility of linguistic features to change over time \cite{Nic1992,Mas2004,Par2008,WicHol2009,GreEtal2010,Ded2011,DedCys2013,Wic2015,GreEtal2017,MurYam2018}. In this tradition, susceptibility to change is evaluated in terms of linguistic stability, which is generally understood as resistance to endogenous change---that is, resistance to mutation in vertical transmission, to the exclusion of horizontal effects. Consider two proto-languages $L$ and $L'$ at a given point in historical time, such that $L$ possesses feature $F$ whilst $L'$ lacks this feature. After a suitable period of time, if all the descendants of $L$ possess feature $F$, and all the descendants of $L'$ lack it, then $F$ is said to display maximal stability over this time period. Conversely, $F$ is said to display maximal instability over this time period if it is found that any individual descendant of $L'$ has exactly the same probability of possessing feature $F$ as any individual descendant of $L$. This ideal scenario assumes that the only forces acting on $L$, $L'$, and their descendants pertain to intergenerational transmission, so that there are no horizontal effects of language contact.}

\newtext{In this light, the tradition of linguistic research described above sees it as a key task to devise methods of stability estimation that can effectively control for the role of horizontal contact in the evolutionary dynamics of language, recovering the vertical signal as cleanly as possible. Some approaches within this tradition rely solely on phylogenetic information, i.e.~information about the distribution of linguistic features among groups of related and unrelated languages \cite{Mas2004,Ded2011}, whilst others combine phylogenetic and areal information \cite{Par2008,WicHol2009,MurYam2018}. In general, however, these approaches seek to control for horizontal effects in an effort to isolate stability in the vertical dimension. For convenience, therefore, we may refer to this tradition as `the stability programme' or `the vertical programme'.}

\newtext{One complication facing the vertical programme is that the actual dynamics of the cultural evolution of language do exhibit extensive horizontal effects; idealized vertical stability is not always recoverable from the signal.} Attested situations of horizontal transfer, in which features are borrowed from one language family to another, range from the multilingualism and diglossia that characterize the linguistic landscapes of major cosmopolitan centres, to more intricate situations of language contact between two or more geographically contiguous language communities \cite{ThoKau1988}. The problem for phylogenetic stability estimation methods arises in these situations from the fact that some linguistic features (e.g.~inflectional markers) are known to be more resistant to horizontal transfer than others \cite{GarArkAmi2014}, while some (e.g.~case systems) are highly vulnerable to simplification in contact situations involving large numbers of second-language learners \cite{BenWin2013}: there is a complex dependence of the rate of horizontally-motivated change on both the type of contact situation, and the nature of the feature itself. Enriching the phylogenetic analysis with areally defined groupings \cite{Par2008,MurYam2018} is only a partial solution, however, as no agreed methods exist for delimiting linguistic areas.

\newtext{A related contrast exists in the field of population genetics between phylogeographic approaches and methods that rely on mathematical models and summary statistics (key quantities summarizing an observation) to infer properties of genetic evolution \cite{Hey2003}. Noting this fact, in this article we pursue a model-based approach to the dynamics of language change. We treat the cultural evolution of language as a combination of two stochastic processes, one operating in the vertical domain, the other operating in the horizontal domain. From this model, we derive a quantity, which we call \emph{(linguistic) temperature}, which expresses the global ratio of unfaithful transmission in both the vertical and horizontal dimensions (mutation) to faithful transmission. By this definition, temperature is different from stability as defined within the vertical programme. Nonetheless, we expect, and indeed demonstrate below, that temperature estimates will be strongly correlated with the stability estimates generated by phylogenetic methods. The expectation of such correlations builds on conceptual considerations which suggest that the mechanisms of endogenous language change (i.e.~the mechanisms that cause mutations in the vertical dimension of intergenerational transmission) are in fact not entirely independent from the mechanisms of contact-driven language change (i.e.~the mechanisms that cause mutations in the horizontal dimension of language contact).}

\newtext{As we show in detail below, the temperature of a linguistic feature can be recovered using this model if two empirically measurable `summary statistics' about that feature are known: the feature's overall frequency across a sample of languages, and a measure of how clumped or scattered the feature is in geographical space. The consequence is that linguistic temperature, a dimensionless measure of the propensity of a linguistic feature to undergo change, is recoverable from synchronic information about that feature's empirical geospatial distributions, without recourse to information about its phylogeny.}

\section*{Results}

\subsection*{Modelling the stochastic dynamics of language}

\newtext{The transmission of a linguistic feature can be faithful or unfaithful whether it takes place on the vertical dimension (i.e.~intergenerationally) or on the horizontal dimension (i.e.~through language contact). As above, faithful transmission in the vertical dimension results in historical stability (in the technical sense of the existing literature), whereas unfaithful transmission in the vertical dimension amounts to endogenous change. On the horizontal dimension, however, transmission can also be either faithful or unfaithful. Faithful horizontal transmission results in simple transfer between languages by contact, often called borrowing \cite{ThoKau1988}. In contrast, unfaithful horizontal transmission occurs when adult learners of a second language (L2) incorporate into their first language (L1) a modified---typically, simplified---version of a feature of L2. This sort of simplification by L2 learners is widely thought to underlie phenomena such as the emergence of impoverished inflectional systems from contact between languages with rich but heterogeneous inflections \cite{BenWin2013}.}

\begin{figure}
  \centering
  \includegraphics[width=0.90\textwidth]{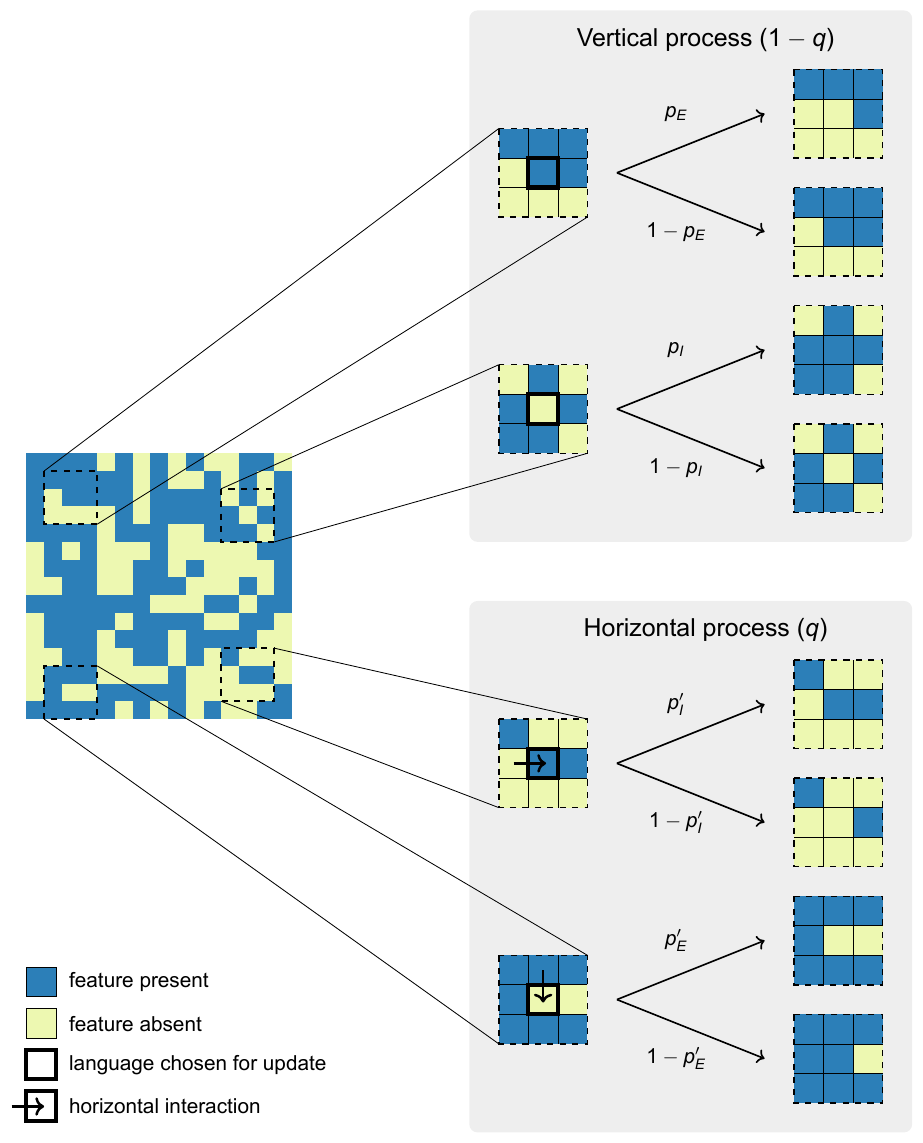}
  \caption{\newtext{{\bf Illustration of the model dynamics.} At each iteration, a random cell of the lattice is chosen for update. A randomly selected feature then undergoes either a horizontal event (with probability $q$) or a vertical event (with probability $1-q$). The value of the feature may flip (from `absent' to `present', or vice versa) due to ingress or egress in either type of event. In a horizontal event, the donor language is drawn randomly from among the focal language's four Von Neumann neighbours.}}\label{fig:model-summary}
\end{figure}

\newtext{To model these interactions, we expand upon an early but under-exploited paradigm in dynamic linguistic typology which proposed to model the dynamics of language as a Markov process in the vertical domain \cite{Gre1978}. This work also suggested, albeit without putting forward a mathematically explicit model of spatial diffusion, that features attesting different ingress and egress rates ought to pattern differently geographically. We here make these assumptions concrete by implementing languages on a spatial substrate; a similar approach, based on computer simulations of a more complex dynamics, has been pursued in \cite{WicHol2009}.} For the sake of mathematical tractability, we assume languages to be distributed on a regular square lattice and each feature to be binary (absent or present) in any given language. Each language on the lattice is subject to a vertical and a horizontal process with respect to each of its features; in our stylized model we assume that each feature evolves independently of the other features. \newtext{The model has five free parameters per feature, $p_I$, $p_E$, $p_I'$, $p_E'$ and $q$, each a probability. In the vertical process, transmission errors occur at rates $p_I$ and $p_E$, where $p_I$ is the probability of innovating a feature which the language lacks (called the feature's vertical ingress rate), and $p_E$ is the probability of losing an existing feature (called the feature's vertical egress rate). In the horizontal process, a feature (or its absence) is copied into the language from one of its immediate neighbours on the lattice. This horizontal transfer is subject to errors as follows: a feature's absence is incorrectly copied as its presence with probability $p_I'$ (horizontal ingress rate), while its presence is incorrectly copied as its absence with probability $p_E'$ (horizontal egress rate). A fifth parameter, $q$, supplies the relative rate of horizontal over vertical events (see Figure \ref{fig:model-summary} for a summary illustration, and Materials and Methods for a complete algorithmic specification of the model).}

\newtext{Similar building blocks (copying, diffusion and mutation) have been used in other fields, for example in modelling the dynamics of opinions in the social sciences \cite{CasForLor2009}. Notably, in biology, models of this general type are often a key component of genetic inference based on summary statistics \cite{Hey2003}. So-called `stepping stone', `island' or `voter' models involve combinations of copying processes and mutation \cite{KimWei1964,CliSud1973,Lig1997,CasForLor2009}, with some differences in the details of the implementation of these components. In some models, each spatial node is occupied by multiple agents \cite{KimWei1964}, whilst in others each node hosts one agent only, similar to our model set-up \cite{CliSud1973}. The model we employ can perhaps best be described as an `asymmetric noisy two-state voter model' \cite{GRANOVSKY1995}. Its behaviour, like that of related models, can be studied using methods and concepts from statistical physics \cite{Oli2003,CasForLor2009,KraRedBen2010,Korolev2010}. In particular, key quantities describing the stationary state can be computed. In our case these are properties of the distribution of features across the lattice. Our approach is therefore similar in spirit to work in population genetics focusing on the inference of parameters of evolutionary processes from summary statistics of observed patterns of genetic diversity, using analytical solutions of stylized models of evolution (for example via coalescence theory) \cite{Kingman1982, Wilkins2002, Hey2003, Wakeley2009, Korolev2010}.}

\begin{figure}
  \centering
  \includegraphics[width=0.8\textwidth]{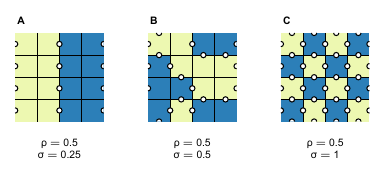}
  \caption{{\bf Feature frequency and isogloss density.} In each of these schematic illustrations, the feature frequency is $\rho = 0.5$ (half of the sites are yellow, the other half are blue). However, isogloss density $\sigma$, defined as the proportion of disagreeing lattice interfaces (white dots), depends on the spatial distribution of the feature. It is low when a feature is present throughout extended domains (\textbf{A}), intermediate when a feature is randomly distributed (\textbf{B}), and large when a feature is scattered (\textbf{C}).}\label{fig:lattice}
\end{figure}

In our lattice model, the statistical properties of the stationary distribution of a linguistic feature depend on the feature's parameters \newtext{$p_I$, $p_E$, $p_I'$, $p_E'$ and $q$}. Useful information about the stationary distribution is contained in two quantities, illustrated in Fig.~\ref{fig:lattice}: the \emph{frequency} $\rho$ with which a particular feature is present across the lattice, and the feature's associated \emph{isogloss density} $\sigma$. The latter quantity is defined as the probability of finding a dialect boundary (an isogloss) between two neighbouring languages such that the feature is found on one side of the boundary but not on the other; similar quantities are sometimes also found as the `density of active interfaces' or `active bonds' (see \cite{CasForLor2009} and references therein). \newtext{Our model is more stylized than that in \cite{WicHol2009}, for instance, and as a consequence of this austere setup, values of $\rho$ and $\sigma$ in the stationary distribution can be calculated analytically in the special case $p_I' = p_E'$, i.e.~when errors in the horizontal transmission of language are not directed. This calculation follows well-established principles in statistical physics \cite{CasForLor2009,KraRedBen2010}, in particular the procedure in \cite{Oli2003}. The following results are a generalization of the analytical solution, verified in numerical simulations for the general case in which $p_I'$ and $p_E'$ are independent. We refer the reader to the Supplementary Material for the analytical derivations and a full description of the numerical simulations.}

\begin{figure}
  \centering
  \includegraphics[width=0.89\textwidth]{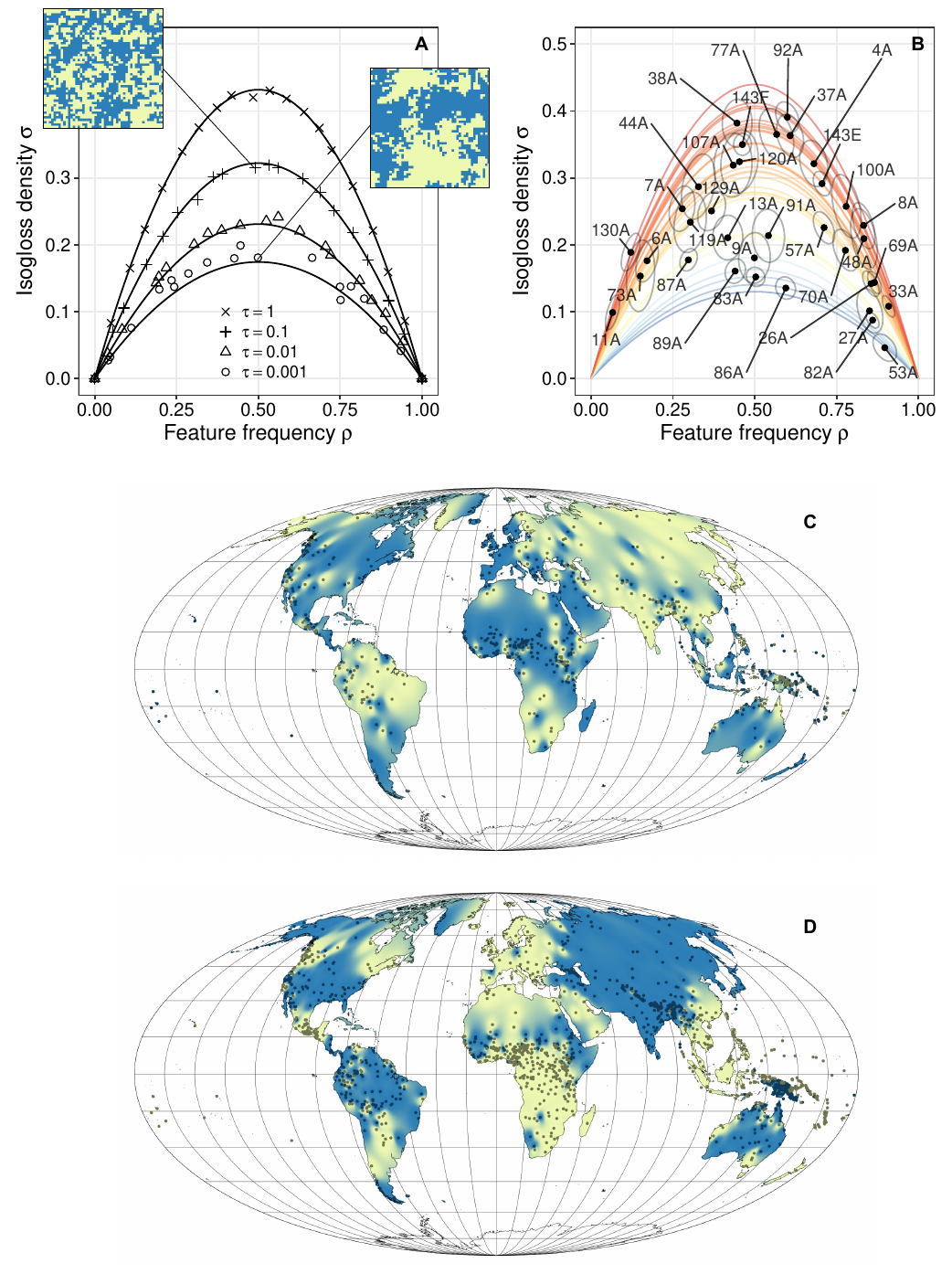}
  \caption{{\bf Statistical properties of the model and empirical measures of feature frequency and isogloss density.} \textbf{(A)} At long times the state of the lattice is characterized by the two quantities feature frequency $\rho$ and isogloss density $\sigma$. We show computer simulations (markers) and analytical solution (curves) for different values of $\tau$. Simulation snapshots of the lattice are shown for two different values of $\tau$. \textbf{(B)} Empirical measurements of feature frequency $\rho$ and isogloss density $\sigma$ for 35 linguistic features, identified by their WALS feature IDs \newtext{(see Table \ref{tab:features-mainpaper} for a key), with 95\% confidence ellipses from the empirical bootstrap.} \textbf{(C)} Empirical geospatial distribution of WALS feature 37A (definite marker). \textbf{(D)} Empirical geospatial distribution of WALS feature 83A (OV word order). Shown are both individual empirical data points (languages, as given by WALS coordinates) and a spatial interpolation (inverse distance weighting) on these points. Blue: feature present, yellow: feature absent. Map projection: \newtext{Mollweide equal-area.}}\label{fig:mainfig}
\end{figure}

The frequency of a feature in the stationary distribution is given by
\newtext{\begin{equation}\label{eq:A:steady-rho}
  \rho = \frac{(1-q)p_I + q p_I'}{(1-q)p + qp'}
\end{equation}
where $p = p_I + p_E$ and $p' = p_I' + p_E'$ represent the total error rates of the vertical and horizontal processes, respectively.} For the isogloss density, we find
\begin{equation}\label{eq:A:steady-sigma}
  \sigma = 2H(\tau) \rho (1- \rho),
\end{equation}
with
\begin{equation}\label{eq:A:Htau}
  H(\tau) = \frac{\pi (1+\tau)}{2K\left(\frac{1}{1+\tau}\right)} - \tau,
\end{equation}
and 
\newtext{
\begin{equation}\label{eq:A:tau}
  \tau = \frac{(1-q)p + qp'}{q(1 - p')}.
\end{equation}
}
The function $K(\cdot)$ denotes the complete elliptic integral of the first kind. From Eq.~(\ref{eq:A:steady-sigma}), the stationary-state isogloss density $\sigma$ is found to be a parabolic function of the feature's overall frequency $\rho$. The height of this parabola is controlled by $H(\tau)$, and hence ultimately by the parameter $\tau$ (Fig.~\ref{fig:mainfig}A). \newtext{This parameter gives the relative rate of unfaithful transmission events (i.e., mutations) over faithful transmission events (Eq.~\ref{eq:A:tau}), and can thus be interpreted as a dimensionless measure of the propensity of the feature to undergo change: lower values of $\tau$ signify a relatively infrequently changing feature, while higher values indicate relative rapidity in change.\footnote{\newtext{In Eq.~(\ref{eq:A:tau}), the denominator does not include a term for faithful transmission events in the vertical process. This may be puzzling at first, but becomes more natural when one realizes that faithful vertical events never change the state of the lattice. These events promote neither order nor disorder, and temperature as an overall measure of disorder is hence not affected by the background of faithful vertical transmission.}} We refer to $\tau$ as \emph{temperature}, and note that, as a dimensionless ratio, it is not calibrated to an overall frequency of transmission events in language. Such a calibration is unnecessary for our purposes, as we are only interested in the relative ranking of different features in terms of $\tau$.}

\subsection*{Inferring linguistic temperatures from geospatial distributions}

\newtext{To arrive at empirical estimates of temperatures of linguistic features, data are then needed from which feature frequencies ($\rho$) and isogloss densities ($\sigma$) can be measured.} Such data are available in the World Atlas of Language Structures or WALS \cite{WALS}, a large-scale typological database also containing spatial information in the form of geographical coordinates for languages. We curated 35 binary or binarized features from the WALS, each of which is recorded for at least 300 languages in the atlas (for full details, see Materials and Methods). For each feature, the frequency $\rho$ is given by the proportion of languages in which that feature is present (rather than absent) in the feature's WALS language sample. \newtext{Isogloss density $\sigma$ was estimated using the 10 geographically nearest neighbour languages of each language in the sample. Finally, the analysis was repeated 1,000 times, resampling languages with replacement in order to generate bootstrap confidence intervals. The results are summarized in Fig.~\ref{fig:mainfig}B, which supplies median feature frequency $\rho$ and isogloss density $\sigma$ for each of the 35 features, together with 95\% bootstrap confidence ellipses. Fig.~\ref{fig:mainfig}C--D provide a detailed illustration of the geospatial distribution of two features, definite marker (WALS feature 37A) and order of object and verb (WALS feature 83A).}

\begin{figure}[t]
  \centering
  \includegraphics[width=0.99\textwidth]{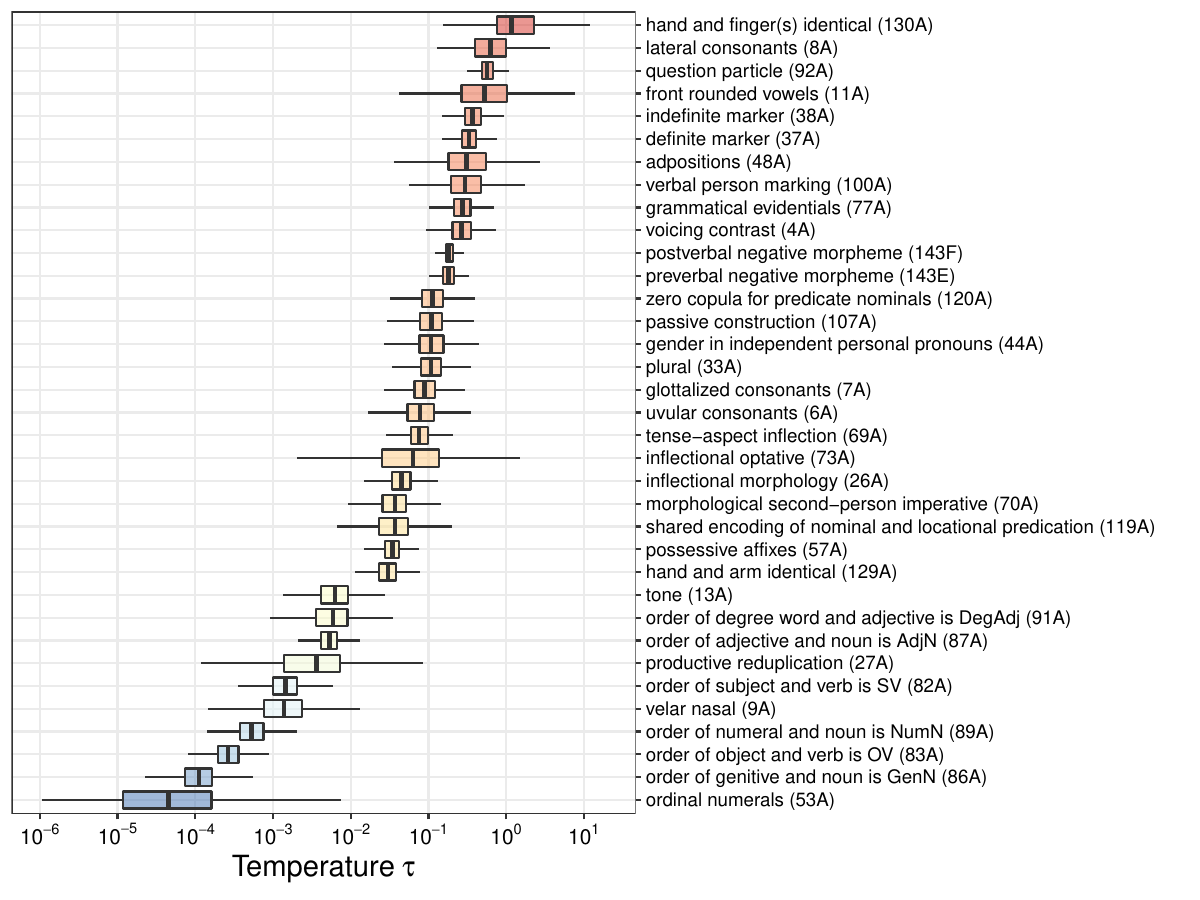}
  \caption{\newtext{{\bf Temperature estimates for the 35 WALS features considered in this study.} The boxplots show the bootstrap distributions over 1,000 runs; central bars represent medians.}}\label{fig:temperatures}
\end{figure}

For a given feature frequency $\rho$, the isogloss density $\sigma$ is fixed by the value of $H(\tau)$ (Eq.~\ref{eq:A:steady-sigma}); this quantity itself is an increasing function of $\tau$ (Eq.~\ref{eq:A:Htau}). Since each of our 35 empirical features lies on a unique parabola in the space spanned by $\rho$ and $\sigma$ (Fig.~\ref{fig:mainfig}), estimating its temperature is now a matter of inverting the function $H(\tau)$. Although the elliptic integral in Eq.~(\ref{eq:A:Htau}) cannot be expressed in terms of elementary functions and $H(\tau)$ thus cannot be inverted analytically, the inversion can be performed numerically \newtext{(see Materials and Methods)}. Using this procedure we obtain an estimate of $\tau$ for any feature for which empirical measurements of frequency $\rho$ and isogloss density $\sigma$ exist. \newtext{Figure \ref{fig:temperatures} supplies the bootstrap distributions of these estimates (for numerical values of the medians, see Table \ref{tab:features-mainpaper}). Estimated temperatures span a range of roughly five orders of magnitude, with word order features tending to have the lowest temperatures and certain lexical, phonological and morphological features the highest.}

\subsection*{Comparison with a phylogenetic method}

\newtext{We have predicted on conceptual grounds that our estimates of linguistic temperature ($\tau$) will be correlated with estimates of vertical stability. To test this prediction, we choose a method of vertical stability estimation that is diametrically opposed to our own spatial procedure of temperature estimation: namely, a method that operates on phylogenetic data to the exclusion of any spatial signal.} Dediu \cite{Ded2011} recovered estimates of the rate of evolution of a selection of linguistic features using two different Bayesian phylogenetic methods and drawing data from two sources---WALS and Ethnologue \cite{Ethnologue}---to control for possible implementation effects. The aggregate rate estimates from this analysis are expressed as the additive inverse of the first component of a principal component analysis (PC1) on the evolution rate rank predicted by each combination of phylogenetic algorithm and dataset. In practice, the higher the PC1 value, the higher the evolution rate of the feature and consequently the lower its stability.

\begin{figure}
  \centering
  \includegraphics[width=0.99\textwidth]{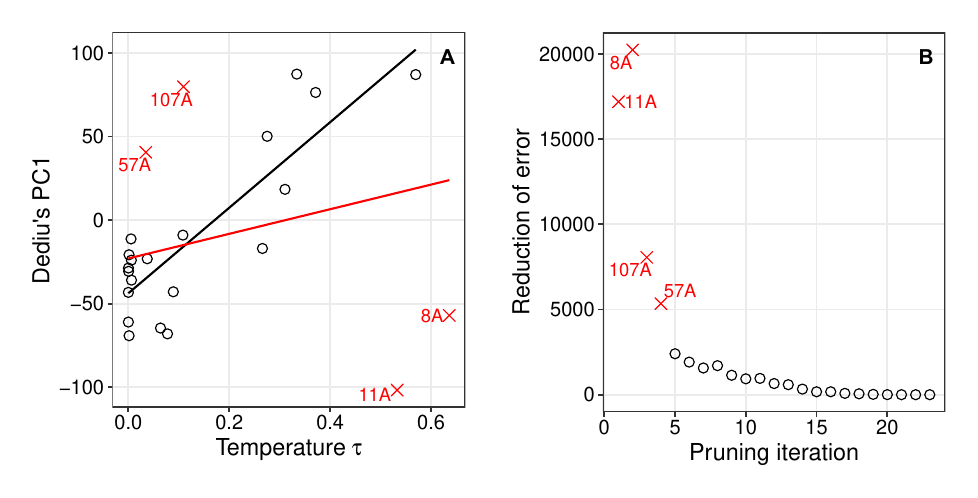}
  \caption{{\bf Regression of our temperature estimates ($\tau$) against Dediu's PC1 \cite{Ded2011}.} \textbf{(A)} Red line: regression with all 24 data points (\newtext{Spearman's $r_S = 0.37$, $p = 0.08$}). Black line: regression with four outliers (red crosses) removed (\newtext{Spearman's $r_S = 0.68$, $p < 0.01$}). \textbf{(B)} Outliers were detected by pruning the dataset recursively for those data points that contributed most to the regression error, quantified as the sum of squared residuals. This identified features 11A, 8A, 107A and 57A as outliers (see text).}\label{fig:comparison}
\end{figure}

In Fig.~\ref{fig:comparison}, we consider the 24 features which fall in the intersection of our list of 35 features and Dediu's list. Regressing our (median) estimates for $\tau$ against Dediu's PC1 (red regression line), we find no evidence of a correlation between the estimates predicted by the two methods (\newtext{Spearman's $r_S = 0.37$, $p = 0.08$}). A number of features, however, are clearly outliers of the regression. To detect these outliers more objectively, we pruned the regression recursively by removing those data points that contributed the greatest error in terms of sum of squared residuals. This procedure classified as outliers the following WALS features: 11A (front rounded vowels), 8A (lateral consonants), 107A (passive construction) and 57A (possessive affixes). Regressing the pruned dataset (Fig.~\ref{fig:comparison}, black regression line), we find a significant high correlation between our $\tau$ estimates and Dediu's PC1 (\newtext{Spearman's $r_S = 0.68$, $p < 0.01$}).

We suggest that, rather than instantiating a lack of correlation between stability (as understood in the vertical programme) and temperature, these outliers are false positives and false negatives of the purely phylogenetic method of stability estimation in \cite{Ded2011}. This is illustrated by the case of front rounded vowels (WALS feature 11A), i.e.~the presence or absence of the vowels /y/ (e.g.~Finnish \emph{kyy}), /ø/ (German \emph{schön}), /œ/ (French \emph{bœuf}) and /{\footnotesize Œ}/ (Danish \emph{grøn}) in a language's phonology \cite{Wic2015}. Front rounded vowels are one of the most stable features in the genetic analysis\footnote{Front rounded vowels are the fourth most stable feature (out of 86) in Dediu's study (Ref.~\cite{Ded2011}, Table S4) and the second most stable (out of 62) in Dediu and Cysouw's metastudy of traditional methods of stability estimation (Ref.~\cite{DedCys2013}, Table 7).} but one of the highest-temperature features in ours (Figure \ref{fig:temperatures}); we argue that evidence from both language change and language acquisition supports our position. On the one hand, front rounded vowels are frequently innovated: historical fronting of the back rounded vowel /u/ to [y] (with or without subsequent phonemicization to /y/) has been documented in a number of languages \cite{HarEtAl2011}. Additionally, front rounded vowels can arise through the influence of /i/ or /j/ on a neighbouring back rounded vowel \cite{IveSal2003}. On the other hand, front rounded vowels are difficult to acquire in situations of language contact: laboratory experiments have shown that second-language learners whose first language lacks these vowels perceive them as more similar to back vowels than front vowels \cite{StrLevLaw2009}. This perceptual assimilation is mirrored in speech production: productions of /y/ by second-language learners are far less advanced in phonetic space than native speakers' productions, and are indeed often perceived as /u/ by the latter \cite{Roc1995}. The fact that front rounded vowels are readily innovated points to a high ingress rate, while frequent acquisition failure by second-language learners in situations of language contact points to a high egress rate. These facts are inconsistent with the high stability predicted by the phylogenetic method, but consistent with our approach, in which high ingress and high egress imply high temperature (Eq.~\ref{eq:A:tau}).

Similar arguments can be made for the remaining outliers. For instance, all Uralic languages employ possessive affixes (e.g.~Finnish \emph{auto-ni} `my car', \emph{auto-si} `your car', etc.),~and the appearance of this system of possession can be dated back to Pre-Proto-Uralic by standard reconstructive techniques \cite{Jan1982}. Possessive affixes are also old in the unrelated Turkic language family \cite{Erd2004}. There is, then, reason to believe that WALS feature 57A on possessive affixes is a false negative of the purely phylogenetic method in \cite{Ded2011}, which classifies possessive affixes as one of the least stable features (Fig.~\ref{fig:comparison}). \newtext{These conclusions are further supported by the fact that an independent method combining a phylogenetic and an areal signal \cite{WicHol2009} agrees with our temperature-based method on three of the four outliers, classifying---like our method, but unlike the phylogenetic method---WALS features 11A and 8A as unstable and 57A as stable. There is, in other words, reason to believe that focusing on the phylogenetic signal to the complete exclusion of the areal dimension leads to a number of features being misclassified or mis-measured in terms of their stability and temperature.}

\section*{Discussion}

\newtext{Estimating the speed of linguistic change is challenging, essentially, because the signal is poor:} although evolutionary and anthropological evidence suggests that human language in its modern form has existed for at least 100,000 years \cite{Bic2007}, the historical evolution of languages is (necessarily) poorly documented. Such documentation only captures a few thousand years for languages with the best coverage, cannot in principle go beyond the introduction of the first writing systems, and does not exist at all for the majority of the world's languages. The rest of the cultural evolution of human language must be reconstructed based on available data; \newtext{in particular, methods for estimating the temporal stability of features of language have traditionally relied on phylogenetic analysis. In this paper, we shown that treating language dynamics as a stochastic process combining both a vertical and a horizontal dimension naturally leads to the notion of linguistic temperature, a dimensionless measure of the propensity of linguistic features to undergo change. Importantly, temperatures of linguistic features can be readily estimated from purely synchronic information:} all that is needed are estimates of feature frequency and isogloss density from a sufficiently large sample of languages, and inversion of Eq.~(\ref{eq:A:Htau}).

We have offered some evidence in support of our method, in the sense that this method is not liable to some of the false positives and false negatives incurred by some purely phylogenetic methods of stability estimation. Turning now to its limitations, we note that our approach currently only applies to binary features, i.e.~features which are either present in or absent from a language. Most genetic methods do not suffer from this limitation: Dediu's \cite{Ded2011} procedure, in particular, can be applied to polyvalent as well as binary features. Interestingly, however, Dediu finds a correlation between estimates for polyvalent and binary (or binarized) features. This suggests that the resolution at which the values of a linguistic variable are recorded may be a minor issue: after all, any polyvalent classification can be reduced to a hierarchy of binary oppositions by a simple translation procedure. Another limitation of our technique is that it treats the evolution of each individual feature independently. Feature interactions are known to exist, however---for example, a language which places objects before verbs is far more likely to also place adverbs before verbs, rather than after them \cite{Gre1963}. It would in principle be possible to generalize our method to cater for polyvalent features and feature interactions, by extending the lattice model in the direction of the Axelrod model of cultural dissemination \cite{Axe1997}. \newtext{The extent to which the behaviour of such a generalized model can be characterized analytically is however not clear, and temperature estimates may have to be obtained in some other way. Similarly, extending the analysis to multiple summary statistics (beyond feature frequency and isogloss density) is likely to lead to analytical challenges, and may necessitate computational inference approaches. Approximate Bayesian computation, for example, is successfully used in population genetics \cite{Beaumont2025}, and has recently been applied to a comparison of genetic and linguistic evolution \cite{AmorimEtAl2013}. Other avenues for extending the present model include exploration of transient long-range geographical connections in addition to local spatial effects, incorporation of more realistic selection and mutation dynamics in both the vertical and the horizontal dimension \cite{BaxEtal2006,Bur2017}, and incorporation of a model of linguistic speciation and a treatment of the resulting geospatial distributions of entire families of languages \cite{Wic2017}.}

\newtext{The derivation of linguistic temperature, together with the empirical demonstration that temperatures of linguistic features are measurable from typological atlases, suggests the existence of large-scale regularities in the transmission of language, in both the vertical and the horizontal dimension. Although the evolutionary trajectories of individual languages are to a large extent moulded by contingencies of history, when the representation of structural features of language are explored at the level of aggregates of languages, regularities emerge. The estimation of linguistic temperatures is but one possible application resulting from work that combines the mathematical analysis of stochastic systems with modern large-scale linguistic datasets, and we expect similar approaches also to be possible in other areas of cultural evolution outside the narrow domain of language.}

\section*{Materials and Methods}

\newtext{
\subsection*{Model}

The model assumes languages to be distributed among the sites or cells of a regular square lattice, and is characterized by five parameters per feature, each a probability: ingress and egress rates in the vertical process ($p_I$ and $p_E$), ingress and egress rates in the horizontal process ($p_I'$ and $p_E'$), and the relative rate of horizontal versus vertical events ($q$). The model is iterated as follows until statistical equilibrium is reached:
\begin{enumerate}
  \item[1.] Initialize the lattice in a random state (for each feature $F$ and lattice cell $C$, $F$ is present in $C$ with probability $0.5$).
\item[2.] Pick a random cell (language) $C$ and a random feature $F$.
\item[3.] Execute one of the following steps:
\begin{enumerate}
\item[3a.] With probability $1 - q$, conduct a vertical event. If $F$ is absent from $C$, acquire $F$ with probability $p_I$ (ingress); if $F$ is present in $C$, lose $F$ with probability $p_E$ (egress).
\item[3b.] With probability $q$, conduct a horizontal event. Pick a random nearest neighbour $C'$ of $C$. If $F$ is absent from $C'$, copy the absence of $F$ to $C$ with probability $1-p_I'$, otherwise set the state of $F$ in $C$ to `present'. If $F$ is present in $C'$, copy its presence to $C$ with probability $1-p_E'$, otherwise set the state of $F$ in $C$ to `absent'.
\end{enumerate}
\item[4.] Go to 2.
\end{enumerate}
The stationary distribution of this model may be studied analytically in the special case $p'_I = p'_E$, and using numerical simulations in the general case, as detailed in the accompanying Supplementary Materials. More realistic spatial substrates can also be considered in numerical simulations, again as outlined in the Supplementary Materials. 
}

\subsection*{Empirical estimation of temperatures}

To estimate empirical temperatures, the latest (2014) version of the WALS Online database \cite{WALS} was downloaded and used as the empirical basis for measures of feature frequency $\rho$ and isogloss density $\sigma$. Since WALS employs a polyvalent coding for most features, a manual pass through the data was first made, retaining only those features that are binary or binarizable on uncontroversial linguistic grounds. Features with fewer than 300 languages in their language sample were discarded to ensure statistically robust results. \newtext{Sign languages were excluded.} This procedure resulted in 35 binary features (see Table \ref{tab:features-mainpaper} for a listing, and Supplementary Materials for detailed information about our binarization scheme).

\begin{table}
  \footnotesize
  \centering
  \caption{\newtext{The 35 linguistic features consulted in this study, ranked in order of decreasing estimated temperature ($\rho$: frequency of feature, $\sigma$: isogloss density, $\tau$: temperature; median values across 1,000 bootstrap samples).}}\label{tab:features-mainpaper}
  \newtext{\begin{tabular}{rllrrr}
Rank & Feature & WALS ID & $\rho$ & $\sigma$ & $\tau$ \\
\hline
1. & hand and finger(s) identical & 130A & 0.12142 & 0.18876 & 1.21789 \\
2. & lateral consonants & 8A & 0.83245 & 0.22907 & 0.63618 \\
3. & question particle & 92A & 0.59955 & 0.39077 & 0.56961 \\
4. & front rounded vowels & 11A & 0.06584 & 0.09890 & 0.53281 \\
5. & indefinite marker & 38A & 0.44569 & 0.38193 & 0.37116 \\
6. & definite marker & 37A & 0.60806 & 0.36313 & 0.33385 \\
7. & adpositions & 48A & 0.83333 & 0.20930 & 0.31049 \\
8. & verbal person marking & 100A & 0.77895 & 0.25744 & 0.29686 \\
9. & grammatical evidentials & 77A & 0.56699 & 0.36507 & 0.27513 \\
10. & voicing contrast & 4A & 0.68078 & 0.32122 & 0.26579 \\
11. & postverbal negative morpheme & 143F & 0.46224 & 0.34988 & 0.18558 \\
12. & preverbal negative morpheme & 143E & 0.70544 & 0.29141 & 0.18094 \\
13. & zero copula for predicate nominals & 120A & 0.45337 & 0.32420 & 0.11337 \\
14. & passive construction & 107A & 0.43432 & 0.31892 & 0.10915 \\
15. & gender in independent personal pronouns & 44A & 0.32804 & 0.28686 & 0.10814 \\
16. & plural & 33A & 0.90901 & 0.10821 & 0.10752 \\
17. & glottalized consonants & 7A & 0.27866 & 0.25397 & 0.08912 \\
18. & uvular consonants & 6A & 0.17108 & 0.17617 & 0.07762 \\
19. & tense-aspect inflection & 69A & 0.86561 & 0.14372 & 0.07507 \\
20. & inflectional optative & 73A & 0.15047 & 0.15337 & 0.06375 \\
21. & inflectional morphology & 26A & 0.85552 & 0.14203 & 0.04492 \\
22. & morphological second-person imperative & 70A & 0.77697 & 0.19176 & 0.03745 \\
23. & shared encoding of nominal and locational predication & 119A & 0.30311 & 0.23385 & 0.03724 \\
24. & possessive affixes & 57A & 0.71175 & 0.22579 & 0.03463 \\
25. & hand and arm identical & 129A & 0.36791 & 0.25061 & 0.03016 \\
26. & tone & 13A & 0.41746 & 0.21060 & 0.00626 \\
27. & order of degree word and adjective is DegAdj & 91A & 0.54177 & 0.21376 & 0.00595 \\
28. & order of adjective and noun is AdjN & 87A & 0.29736 & 0.17745 & 0.00533 \\
29. & productive reduplication & 27A & 0.85054 & 0.10143 & 0.00361 \\
30. & order of subject and verb is SV & 82A & 0.86013 & 0.08752 & 0.00145 \\
31. & velar nasal & 9A & 0.49893 & 0.18057 & 0.00139 \\
32. & order of numeral and noun is NumN & 89A & 0.44015 & 0.16075 & 0.00053 \\
33. & order of object and verb is OV & 83A & 0.50317 & 0.15203 & 0.00027 \\
34. & order of genitive and noun is GenN & 86A & 0.59497 & 0.13545 & 0.00011 \\
35. & ordinal numerals & 53A & 0.89720 & 0.04624 & 0.00005 \\
\hline
\end{tabular}
}
\end{table}

\newtext{Temperatures $\tau$ were estimated by taking 1,000 bootstrap samples for each feature from the feature's WALS language sample. Feature frequency $\rho$ is the proportion of languages attesting the feature in that feature's bootstrap sample. In the calculation of isogloss densities $\sigma$, the 10 geographically nearest neighbours of each language were considered, found using the haversine formula assuming a perfectly spherical earth. For each language--language pair, an isogloss was recorded if the two languages differed in their value for the feature in question; isogloss density is the number of isoglosses divided by the total number of pairs. Variation in the number of nearest neighbours considered did not have an effect on the results (see Supplementary Materials). Temperatures were recovered by inverting Eq.~(\ref{eq:A:Htau}) using a computationally generated hash table, the complete elliptic integral solved numerically using the arithmetic--geometric mean.

To ensure that our method is catching a universal signal about temperatures of linguistic features, rather than contingent properties of particular geographical areas, we performed the two hemispheres test described in \cite{WicHol2009}. In this test, the analysis is carried out for the Western and Eastern hemispheres separately, and the temperature estimates arising from the two analyses correlated. As detailed in the Supplementary Materials, the Spearman correlation for temperature estimates in the two hemispheres was found to be $0.52$, statistically significant at $p < 0.01$. This compares with the Spearman correlation reported earlier for a genealogical--areal method of stability estimation in the same test, $0.51$ \cite{WicHol2009}.}

\subsection*{Comparison with phylogenetic method}

For the comparison with the phylogenetic method, Table S1 in the Supplementary Material to \cite{Ded2011} were consulted and only those features were selected for comparison for which our binarization schemes agreed; the PC1 values for the intersecting features were then gathered from Table S4 of that publication.

\section*{Data and code availability}

This article reports no original data; the data we used are freely available from \url{https://wals.info/}. Data analysis and simulation code can be downloaded from \url{https://github.com/hkauhanen/littlesquares/}.

\section*{Competing interests}

We have no competing interests.

\section*{Author contributions}

\newtext{HK, DG, TG and RB-O designed the study, defined the model, analysed the data and wrote the manuscript together. HK and DG produced the visualizations. HK wrote the data analysis and simulation codes. TG derived the analytical solution of the lattice model, with inputs from HK and DG. All authors gave final approval for publication.}

\section*{Acknowledgements}

\newtext{We thank Dan Dediu, Danna Gifford, George Walkden and Jon F.~Wilkins for comments.}

\section*{Funding}

\newtext{HK was funded by Emil Aaltonen Foundation, The Ella and Georg Ehrnrooth Foundation, the Economic and Social Research Council (ES/S011382/1), and the Federal Ministry of Education and Research (BMBF) and the Baden-Württemberg Ministry of Science as part of the Excellence Strategy of the German Federal and State Governments. TG acknowledges funding from the Spanish Ministry of Science, Innovation and Universities, the Agency AEI and FEDER (EU) under the grant PACSS (RTI2018-093732-B-C22), and the Maria de Maeztu program for Units of Excellence in R\&D (MDM-2017-0711).}

\nocite{Mor1971}
\nocite{Fer2016}
\bibliography{tau}
\bibliographystyle{Science_mod}

\end{document}


\author[1]{Henri Kauhanen\thanks{Corresponding author: henri.kauhanen@uni-konstanz.de}}
\author[2]{Deepthi Gopal} 
\author[3,4]{Tobias Galla}
\author[5]{Ricardo Berm\'udez-Otero}
\affil[1]{Zukunftskolleg, University of Konstanz, Universit\"atsstra{\ss}e 10, 78464 Konstanz, Germany}
\affil[2]{Department of Theoretical and Applied Linguistics, University of Cambridge, Sidgwick Avenue, Cambridge, CB3 9DA, UK}
\affil[3]{Department of Physics and Astronomy, School of Natural Sciences, The University of Manchester, Oxford Road, Manchester, M13 9PL, UK}
\affil[4]{Instituto de F\'isica  Interdisciplinar y Sistemas  Complejos, IFISC (CSIC-UIB), Campus Universitat Illes Balears,  E-07122 Palma de Mallorca, Spain}
\affil[5]{Department of Linguistics and English Language, The University of Manchester, Oxford Road, Manchester, M13 9PL, UK}

\maketitle

{\footnotesize
\tableofcontents
}

\section{Introduction}

This document provides further technical details of the model, its analytical solution, numerical simulations, and the resultant procedure for estimating temperatures of linguistic features from empirical data.

We begin by addressing a limited special case of the model, the case where $p'_I = p'_E = 0$, i.e.~when the horizontal process is completely faithful. The stationary distribution of this model can be analytically calculated for an isotropic lattice. Details of the calculation are given in Section \ref{app:solution}. We then generalize the model first by assuming $p'_I \neq 0$ and $p'_E \neq 0$ but retaining the requirement that $p'_I = p'_E$. This generalization also turns out to yield to an analytical treatment, as detailed in Section \ref{app:generalization}. In Section \ref{app:full-generality} we drop the requirement that $p'_I = p'_E$. The full analytical solubility is lost for this more general model, but we show, using numerical simulations, that equations (1)--(4) of the main paper, which supply feature frequency, isogloss density and temperature at the stationary distribution, are valid for this general formulation also. Section \ref{app:robustness} reports the results of a number of sanity checks in order to show that our method for evaluating empirical temperatures is robust to (a) choice of geographical area, (b) choice of spatial scale, (c) the spatial structure of the underlying substrate, and (d) potential idiosyncracies arising from the empirical spatial distributions of language families. Finally, Section \ref{sec:wals-details} supplies detailed metadata about the features consulted in our study.

\section{Model: faithful horizontal transmission}\label{app:solution}

To facilitate the analytical exploration of the model's behaviour, we first study the special case $p'_I = p'_E = 0$, i.e.~we assume that all horizontal events result in the faithful transmission of a feature from one lattice node to another.

For the purposes of the analytical calculation we will treat the model as a spin system on a two-dimensional regular square lattice with $N = L\times L$ sites and periodic boundary conditions; comparable approaches are available in the literature on voter models (\citeOli, \citeKraRedBen). We write $s(\vec x) \in \{-1,1\}$ for the spin at lattice site $\vec x = (x_1, x_2)$; $S(\vec x) = \langle s(\vec x)\rangle$ for the average spin of $\vec x$ (over realizations of the stochastic process); and $m = \sum_{\vec x} S(\vec x)/N$ for the mean magnetization over the entire lattice. The feature frequency $\rho$, or fraction of up-spins in the system, is related to $m$ by the identity $\rho = (m+1)/2$. We further write $S(\vec x, \vec y) = \langle s(\vec x)s(\vec y)\rangle$ for the pair correlation of $s(\vec x)$ and $s(\vec y)$. In summations, $\vec x'$ is understood to index the set of von Neumann neighbours of site $\vec x$, i.e.~the set
\begin{equation}
  \{(x_1 - 1, x_2), (x_1 + 1, x_2), (x_1, x_2 - 1), (x_1, x_2 + 1)\}.
\end{equation}

In Sections \ref{sec:calc1}--\ref{sec:calc4} below we calculate analytically the stationary-state feature frequency and isogloss density. While we are not aware of existing work in the exact same model, the calculation follows well established principles from the analytical treatments of voter models in statistical mechanics (see (\citeCasForLor, \citeKraRedBen) and references therein). In particular a very similar calculation is carried out in (\citeOli). The novelty of our work is therefore not in the calculation itself, but in the application of the analytical calculation to define temperature for linguistic features.

\subsection{Spin flip probability}\label{sec:calc1}

Central to our analytical derivations is the spin flip probability, i.e. the probability with which the spin at site $\vec x$ changes its state from $-1$ to $+1$ or vice versa, if it is selected for potential update. In our model this is of the form
\begin{equation}
  w(\vec x) = (1-q) A(\vec x) + q B(\vec x),
\end{equation}
where $A(\vec x)$ is the contribution of the ingress--egress process and $B(\vec x)$ the contribution of the spatial (voter) process. These are
\begin{equation}
  A(\vec x) = \frac{1 - s(\vec x)}{2} p_I + \frac{1 + s(\vec x)}{2} p_E
  = \frac{1}{2} \left[p_I + p_E + (p_E - p_I)s(\vec x)\right]
\end{equation}
and
\begin{equation}
  B(\vec x) = \frac{1}{4}\sum_{\vec x'} \frac{1 - s(\vec x)s(\vec x')}{2}
  = \frac{1}{2} \left[1 - \frac{1}{4} \sum_{\vec x'} s(\vec x)s(\vec x') \right],
\end{equation}
where it is important to remember that the summation over $\vec x'$ runs over the four nearest neighbours of $\vec x$. Hence we have
\begin{equation}\label{eq:spin-flip-probability}
  w(\vec x) = \frac{1-q}{2} \left[p_I + p_E + (p_E - p_I)s(\vec x)\right]
  + \frac{q}{2} \left[1 - \frac{1}{4}\sum_{\vec x'} s(\vec x)s(\vec x') \right].
\end{equation}

\subsection{Stationary-state feature frequency $\rho$}\label{sec:calc2}

The spin at $\vec x$ changes by the amount $-2s(\vec x)$ with probability $\frac{1}{N} w(\vec x)$, the prefactor $1/N$ representing the probability of $\vec x$ being picked for update. Consequently, the mean spin $S(\vec x) = \langle s(\vec x)\rangle$ evolves as
\begin{equation}\label{eq:sdiff}
  S(\vec x, t + \Delta t) - S(\vec x, t) = \frac{1}{N} \langle -2w(\vec x)s(\vec x) \rangle,
\end{equation}
where $\Delta t$ is the time step associated with each attempted spin flip. Bearing in mind that $s(\vec x)s(\vec x) = 1$ and plugging Eq.~(\ref{eq:spin-flip-probability}) in, this implies
\begin{equation}
  \frac{S(\vec x, t + \Delta t) - S(\vec x,t)}{1/N}
  = (1-q) \left[p_I - p_E - (p_I + p_E)S(\vec x) \right]
  + q\left[ -S(\vec x) + \frac{1}{4} \sum_{\vec x'} S(\vec x') \right].
\end{equation}
Taking the sum over all sites $\vec x$, one has
\begin{equation}
  \begin{split}
  \frac{m(t + \Delta t) - m(t)}{1/N}
    &= (1-q)(p_I - p_E) - (1-q)(p_I + p_E) \frac{1}{N} \sum_{\vec x} S(\vec x) - \\
    &\quad\quad - \frac{q}{N} \sum_{\vec x} S(\vec x) + \frac{q}{4N} \sum_{\vec x}\sum_{\vec x'} S(\vec x').
  \end{split}
\end{equation}
Now $\sum_{\vec x}\sum_{\vec x'} S(\vec x') = 4\sum_{\vec x} S(\vec x)$, since the LHS is the sum of the four von Neumann neighbours of all lattice sites, so that each site, having four neighbours, gets counted four times. Using $m = \sum_{\vec x} S(\vec x)/N$, we then have
\begin{equation}
  \frac{m(t + \Delta t) - m(t)}{1/N} = (1-q)\left[p_I - p_E - (p_I + p_E)m\right].
\end{equation}
With the standard choice $\Delta t = 1/N$, and taking the limit $N \to \infty$ (i.e.~the continuous-time limit $\Delta t \to 0$), we thus find
\begin{equation}
  \frac{dm}{dt} = (1-q)\left[p_I - p_E -(p_I + p_E)m\right].
\end{equation}
Hence the mean magnetization in the stationary state ($dm/dt = 0$) is
\begin{equation}\label{eq:emm}
  m = \frac{p_I - p_E}{p_I + p_E}.
\end{equation}
From this, using $\rho = (m+1)/2$, we find
\begin{equation}\label{eq:steady-rho}
  \rho = \frac{p_I}{p_I + p_E}
\end{equation}
for the fraction of up-spins in the stationary state.

\subsection{Pair correlation function}\label{sec:calc3}

To compute the pair correlation $S(\vec x, \vec y) = \langle s(\vec x)s(\vec y) \rangle$, we note that $s(\vec x)s(\vec y)$ changes by the amount $-2s(\vec x)s(\vec y)$ if either $\vec x$ or $\vec y$ flips spin. Assuming $\vec x \neq \vec y$, and working directly in the continuous-time limit, we have
\begin{equation}\label{eq:corr_start}
  \frac{dS(\vec x, \vec y)}{dt} = \langle -2\left[w(\vec x) + w(\vec y)\right]s(\vec x)s(\vec y)\rangle.
\end{equation}
After some algebra we find
\begin{equation}
  \begin{split}
    \frac{d S(\vec x, \vec y)}{dt} &= (1-q) \left[ (p_I - p_E) [S(\vec x) + S(\vec y)] - 2(p_I + p_E)S(\vec x, \vec y) \right] + \\
    &\quad\quad + q \left[ - 2S(\vec x, \vec y) + \frac{1}{4} \sum_{\vec x'} S(\vec x', \vec y) + \frac{1}{4} \sum_{\vec y'} S(\vec x, \vec y') \right],
  \end{split}
 \end{equation}
where the summation over $\vec y'$ is over the four nearest neighbours of $\vec y$.\\

We now assume translation invariance and write $C(\vec r) = C(\vec x - \vec y) = S(\vec x, \vec y)$. Then, for $\vec r \neq \vec 0$,
\begin{equation}
  \begin{split}
    \frac{dC(\vec r)}{dt} &= (1-q) \left[ (p_I - p_E) [S(\vec x) + S(\vec y)] - 2(p_I + p_E) C(\vec r) \right] + \\
    &\quad\quad + q \left[ -2C(\vec r) + \frac{1}{4} \sum_{\vec x'} C(\vec x' - \vec y) + \frac{1}{4}\sum_{\vec y'} C(\vec x - \vec y') \right].
  \end{split}
\end{equation}
Due to translation invariance the two summations on the RHS coincide, and we have (always restricting $\vec r \neq \vec 0$)
\begin{equation}\label{eq:pair-correlation-function}
  \frac{dC(\vec r)}{dt} = (1-q) \left[ (p_I - p_E) [S(\vec x) + S(\vec y)] - 2(p_I + p_E)C(\vec r) \right] + 2q\Delta C(\vec r),
\end{equation}
where $\Delta$ is the lattice Laplace operator
\begin{equation}
  \Delta f(\vec x) = -f(\vec x) + \frac{1}{4}\sum_{\vec x'} f(\vec x').
\end{equation}
We note that we always have the boundary condition $C(\vec 0, t) = 1$ for all $t$ (self-correlation is $1$ at all times, as $s(\vec x)^2=1$).

\subsection{Stationary-state isogloss density $\sigma$}\label{sec:calc4}

If $C(\vec e_1 ,t)$ were known, where $\vec e_1$ is the unit vector $(1,0)$, the isogloss density could be obtained via the identity
\begin{equation}
  \sigma (t) = \frac{1 - C(\vec e_1 ,t)}{2}.
\end{equation}
Thus, knowing the limiting ($t \to \infty$) value of $C(\vec e_1)$ would imply the stationary-state isogloss density $\sigma$.

At the stationary state, $S(\vec x) + S(\vec y) = 2m$ and $\frac{dC(\vec r)}{dt} = 0$. Assuming $\vec r \neq \vec 0$, Eq.~(\ref{eq:pair-correlation-function}) then implies
\begin{equation}
  0 = (1-q) \left[ 2(p_I - p_E)m - 2(p_I + p_E)C(\vec r) \right] + 2q \Delta C(\vec r),
\end{equation}
in other words,
\begin{equation}\label{eq:before-alpha}
  \begin{split}
    0 &= (1-q) \left[ (p_I - p_E) m - (p_I + p_E) C(\vec r) \right] + q\Delta C(\vec r) \\
    &= (1-q) (p_I + p_E) \left[ \frac{p_I - p_E}{p_I + p_E}m - C(\vec r) \right] + q\Delta C(\vec r) \\
    &= (1-q)(p_I + p_E)\left[ m^2 - C(\vec r) \right] + q \Delta C(\vec r).
  \end{split}
\end{equation}
In the special case $q = 0$ (i.e., no spatial interaction between neighbouring sites), this implies $C(\vec r) = m^2$ for all $\vec r \neq \vec 0$. Thus for any two sites $\vec x \neq \vec y$, $\langle s(\vec x)s(\vec y)\rangle = m^2$, indicating that spins at different sites are fully uncorrelated. This is what one would expect, as all sites operate independently when $q=0$.

In the special case $q = 1$ (no ingress--egress dynamics within the sites), on the other hand, we obtain $\Delta C(\vec r) = 0$ for $\vec r \neq 0$. We also have $C(\vec 0) = 0$. This implies $C(\vec r) = 1$ everywhere, so either all spins are up or all are down. This is the only possible stationary state when the only dynamics is through nearest-neighbour interactions.

Now suppose $0 < q < 1$. Dividing Eq.~(\ref{eq:before-alpha}) by $q$ we obtain
\begin{equation}\label{eq:to-be-solved}
  \begin{split}
    0 &= \frac{1-q}{q}(p_I + p_E) \left[m^2 - C(\vec r)\right] + \Delta C(\vec r) \\
    &= \frac{(q-1)(p_I + p_E)}{q} \left[C(\vec r) - m^2\right] + \Delta C(\vec r) \\
    &= \alpha C(\vec r) - \alpha m^2 + \Delta C(\vec r),
  \end{split}
\end{equation}
where we write
\begin{equation}\label{eq:alpha}
  \alpha = \frac{(q-1)(p_I + p_E)}{q}.
\end{equation}
Now let
\begin{equation}\label{eq:defc}
  c(\vec r) = C(\vec r) - m^2.
\end{equation}
We note that $\Delta C(\vec r)=\Delta c(\vec r)$. To solve Eq.~(\ref{eq:to-be-solved}), it then suffices to solve (for $\vec r \neq \vec 0$)
\begin{equation}\label{eq:little-c}
  \Delta c(\vec r) + \alpha c(\vec r) = 0
\end{equation}
subject to the condition
\begin{equation}\label{eq:little-c-condition}
  c(\vec 0) = 1 - m^2.
\end{equation}

We now first focus on the equation  
\begin{equation}\label{eq:G}
  \Delta G(\vec r) + \alpha G(\vec r) = -\delta_{\vec r,\vec 0},
\end{equation}
 for all $\vec r$ (including $\vec r = \vec 0$), and where $\delta_{\vec x,\vec y}$ is the Kronecker delta.

Let $G_{\alpha}(\vec r)$ be a solution of Eq. (\ref{eq:G}). Then
\begin{equation}\label{eq:little-c-G}
  c(\vec r) = (1-m^2) \frac{G_{\alpha}(\vec r)}{G_{\alpha}(\vec 0)}
\end{equation}
is a solution of Eqs.~(\ref{eq:little-c}) and (\ref{eq:little-c-condition}). This can be seen as follows: first, from Eq.~(\ref{eq:little-c-G}),
\begin{equation}
  c(\vec 0) = (1-m^2)\frac{G_{\alpha}(\vec 0)}{G_{\alpha}(\vec 0)} = 1-m^2,
\end{equation}
so the condition in Eq.~(\ref{eq:little-c-condition}) is met. Second, we need to show that Eqs.~(\ref{eq:G}) and (\ref{eq:little-c-G}) imply $\Delta c(\vec r) + \alpha c(\vec r) = 0$ for $\vec r \neq \vec 0$. For $\vec r \neq \vec 0$ we have
\begin{equation}\label{eq:G-with-r-nonzero}
  \Delta G(\vec r) + \alpha G(\vec r) = 0
\end{equation}
from Eq.~(\ref{eq:G}). The quantity $c(\vec r)$ in Eq.~(\ref{eq:little-c-G}) is proportional to $G_{\alpha}(\vec r)$ with a proportionality constant $(1-m^2)/G_{\alpha}(\vec 0)$ which is independent of $\vec r$. Given that $G_{\alpha}(\vec r)$ fulfills Eq.~(\ref{eq:G-with-r-nonzero}) for $\vec r \neq \vec 0$ it is then clear that $c(\vec r)$ fulfills $\Delta c(\vec r) + \alpha c(\vec r) = 0$ for $\vec r \neq \vec 0$, i.e.~Eq.~(\ref{eq:little-c}).
\\

So we are left with the task of finding a solution of 
\begin{equation}\label{eq:big-GG}
  \Delta G(\vec r) + \alpha G(\vec r) = -\delta_{\vec r,\vec 0}.
\end{equation}
Let us write $G(\vec r)$ in Fourier representation:
\begin{equation}\label{eq:fourier}
G(\vec r)=\frac{1}{(2\pi)^2} \int_{-\pi}^{\pi} \int_{-\pi}^{\pi}dk_1 dk_2~ e^{i\vec k \cdot \vec r} \widehat G(\vec k),
\end{equation}
where $\vec k=(k_1,k_2)$ and $\widehat G(\vec k)$ is the Fourier transform of $G(\vec r)$, i.e.
\begin{equation}
\widehat G(\vec k) = \sum_{\vec r} e^{-i \vec k \cdot \vec r} G(\vec r) = \sum_x \sum_y e^{-i x k_1 -i y k_2} G(x,y).
\end{equation}
Applying this to both sides of Eq.~(\ref{eq:big-GG}) leads to 
\begin{equation}\label{eq:help}
\sum_{x,y} e^{-i x k_1 -i y k_2} \left[\Delta G(\vec r) + \alpha G(\vec r)\right]= - 1.
\end{equation}
Next, notice
\begin{equation}
  \begin{split}
\sum_{x,y} e^{-i x k_1 -i y k_2} \Delta G(x,y ) &= \frac{1}{4}\sum_{x,y} e^{-i x k_1 -i y k_2} \left[G(x+1,y)+G(x-1,y)\right. + \\
&\quad\quad \left. +G(x,y+1)+G(x,y-1)-4G(x,y)\right] \\
 &= \frac{1}{4}\sum_{x,y} \left[e^{-i (x-1) k_1 -i y k_2} +e^{-i (x+1) k_1 -i y k_2} \right. + \\
    &\quad\quad \left. +e^{-i x k_1 -i (y-1) k_2}  + e^{-i x k_1 -i (y+1) k_2} -4 e^{-i x k_1 -i y k_2}\right ] G(x,y) \\
 &= \frac{1}{4}\sum_{x,y} \left[\underbrace{e^{i k_1 } +e^{-i k_1 }}_{=2\cos k_1} +\underbrace{e^{i k_2}  + e^{ -i k_2}}_{=2\cos k_2} -4 \right ] e^{-i x k_1 -i y k_2} G(x,y) \\
 &= \frac{1}{2}\left[ \cos k_1 + \cos k_2  -2 \right] \sum_{x,y} e^{-i x k_1 -i y k_2} G(x,y) \\
 &= \frac{1}{2}\left[ \cos k_1 + \cos k_2  -2 \right] \widehat G(\vec k).
  \end{split}
\end{equation}
Using this in Eq.~(\ref{eq:help}) gives
\begin{equation}
\left[\frac{1}{2}\left( \cos k_1 + \cos k_2  -2 \right )+\alpha\right] \widehat G(\vec k)
=-1, 
\end{equation}
in other words
\begin{equation}
\widehat G(\vec k)=\frac{1}{   1-\alpha-\frac{1}{2}\left[\cos k_1 + \cos k_2\right ] }.
\end{equation}
Using Eq.~(\ref{eq:fourier}) we then have
\begin{equation}
  G_\alpha (\vec r) = \frac{1}{(2\pi)^2} \int_{-\pi}^{\pi} \int_{-\pi}^{\pi} \frac{e^{i\vec k \cdot \vec r} dk_1 dk_2}{1 - \alpha - \frac{1}{2}(\cos k_1 + \cos k_2)}.
\end{equation}
Hence
\begin{equation}
  G_\alpha (\vec 0) = \frac{1}{(2\pi)^2} \int_{-\pi}^{\pi} \int_{-\pi}^{\pi} \frac{dk_1dk_2}{1 - \alpha - \frac{1}{2}(\cos k_1 + \cos k_2)}.
\end{equation}
The integrand is symmetric with respect to $k_1 \leftrightarrow -k_1$ and $k_2 \leftrightarrow -k_2$ respectively, due to the identity $\cos(k)=\cos(-k)$. Hence
\begin{equation}
  \begin{split}
  G_\alpha (\vec 0) 
  &= \frac{1}{\pi^2} \int_0^{\pi} \int_0^{\pi} \frac{dk_1 dk_2}{1 - \alpha - \frac{1}{2}(\cos k_1 + \cos k_2)} \\
  &= \frac{1}{1-\alpha} \frac{1}{\pi^2} \int_0^{\pi} \int_0^{\pi} \frac{dk_1 dk_2}{1 - \frac{1}{2(1-\alpha)}(\cos k_1 + \cos k_2)}.
  \end{split}
\end{equation}
This expression is related to the complete elliptic integral of the first kind (\citeMor). We use the following notation:
\begin{equation}
K(z)=\frac{1}{2\pi}\int_0^\pi du \int_0^\pi dv \frac{1}{1-\frac{z}{2}(\cos u + \cos v)}.
\end{equation}
Hence we have
\begin{equation}\label{eq:gk}
  G_\alpha(\vec 0) = \frac{2}{\pi(1-\alpha)} K\left(\frac{1}{1-\alpha}\right) = \frac{2K_\alpha}{\pi (1-\alpha)},
\end{equation}
where we abbreviate
\begin{equation}
  K_\alpha = K\left( \frac{1}{1-\alpha} \right)
\end{equation}
for convenience.

Next, we write the Laplacian $\Delta G_{\alpha} (\vec 0)$ in full:
\begin{equation}
  \Delta G_{\alpha} (\vec 0) = -G_{\alpha}(\vec 0) + \frac{1}{4} [G_{\alpha}(1,0) + G_{\alpha}(-1,0) + G_{\alpha}(0,1) + G_{\alpha}(0,-1)].
\end{equation}
Assuming isotropy, each term in the square brackets is equal to $G_{\alpha}(\vec e_1) = G_{\alpha} (1,0)$. Hence Eq. (\ref{eq:big-GG}), evaluated at $\vec r=\vec 0$, takes the form
\begin{equation}
  G_{\alpha} (\vec e_1) + (\alpha - 1)G_{\alpha}(\vec 0) = -1,
\end{equation}
in other words
\begin{equation}
  G_{\alpha}(\vec 0) - G_{\alpha}(\vec e_1) = 1 + \alpha G_{\alpha} (\vec 0).
\end{equation}
From this we find
\begin{equation}
  1 - \frac{G_{\alpha}(\vec e_1)}{G_{\alpha}(\vec 0)} = \left[\alpha + \frac{1}{G_{\alpha}(\vec 0)} \right].
\end{equation}
Now using Eqs. (\ref{eq:defc}) and (\ref{eq:little-c-G}), 
\begin{equation}
  C(\vec r) = m^2 + c(\vec r) = m^2 + (1-m^2) \frac{G_{\alpha}(\vec r)}{G_{\alpha}(\vec 0)}.
\end{equation}
The stationary-state isogloss density $\sigma$ is then
\begin{equation}
  \begin{split}
    \sigma &= \frac{1}{2} [1 - C(\vec e_1)] \\
    &= \frac{1}{2} \left[1 - m^2 - (1-m^2) \frac{G_{\alpha}(\vec e_1)}{G_{\alpha} (\vec 0)} \right] \\
    &= \frac{1}{2} (1-m^2) \left[1 - \frac{G_{\alpha}(\vec e_1)}{G_{\alpha}(\vec 0)} \right] \\
    &= \frac{1}{2} (1-m^2) \left[\alpha + \frac{1}{G_{\alpha}(\vec 0)}\right] \\
    &= \frac{1}{2} (1-m^2) \left[\alpha + \frac{\pi(1-\alpha)}{2K_{\alpha}}\right],
  \end{split}
\end{equation}
where we have used Eq. (\ref{eq:gk}).
On the other hand,
\begin{equation}
    1-m^2 = 1 - \frac{(p_I - p_E)^2}{(p_I + p_E)^2}
    = 4\left(\frac{p_I}{p_I + p_E}\right)\left(\frac{p_E}{p_I + p_E}\right)
    = 4\rho (1-\rho),
\end{equation}
so that
\begin{equation}\label{eq:sigmarho}
  \sigma = 2\rho (1-\rho) \left[\alpha + \frac{\pi (1-\alpha)}{2K_{\alpha}} \right].
\end{equation}
Recalling that
\begin{equation}\label{eq:alphatau}
  \alpha = -\tau = - \frac{(1-q)(p_I + p_E)}{q},
\end{equation}
see Eq. (\ref{eq:alpha}), and defining $\tau=-\alpha$, i.e. $K_{\alpha} = K_{-\tau}$, yields our final equation for the stationary-state isogloss density:
\begin{equation}\label{eq:parabola}
  \sigma = 2 H(\tau) \rho (1-\rho),
\end{equation}
where 
\begin{equation}\label{eq:Htau}
  H(\tau) = \frac{\pi (1 + \tau)}{2K\left(\frac{1}{1+\tau}\right)} - \tau.
\end{equation}
The expression in Eq. (\ref{eq:parabola}) is a downward-opening parabola in $\rho$ whose height is fixed by $H(\tau)$, where
\begin{equation}\label{eq:tau}
  \tau = \frac{(1-q)(p_I + p_E)}{q}.
\end{equation}

\subsection{Sanity check: the limits $\tau \to 0$ and $\tau \to \infty$}

The complete elliptic integral $K(z)$ has the following known properties (\citeFer):
\begin{equation}
  \lim_{z\to 0} K(z) = \frac{\pi}{2}
  \quad\textnormal{and}\quad
  \lim_{z\to 1} K(z) = \infty.
\end{equation}

Now consider the limit $\tau \to 0$. This limit is relevant when either $q\to 1$, or both $p_E\to 0$ and $p_I\to 0$, see Eq. (\ref{eq:tau}). These are situations in which the spatial (voter) process dominates the ingress--egress dynamics. In this limit $1/(1+\tau) \to 1$, so that $K_{-\tau} = K(1/(1+\tau)) \to \infty$. Noting that Eq.~(\ref{eq:Htau}) can be written in the form
\begin{equation}\label{eq:Htau-different-form}
  H(\tau) = \tau \left[ \frac{\pi}{2K_{-\tau}} - 1\right] + \frac{\pi}{2K_{-\tau}},
\end{equation}
we then find that $H(\tau) \to 0$. Thus, in the limit where the spatial (voter) process completely overtakes the ingress--egress process, the stationary-state isogloss density is zero, indicating that all sites agree in their spin.

Next consider $\tau \to \infty$. This limit occurs when $p_I+p_E>0$ and $q\to 0$, i.e. the ingress--egress process dominates. Then $1/(1+\tau) \to 0$, so that $K_{-\tau} \to \pi/2$. From Eq.~(\ref{eq:Htau-different-form}), we then find that $H(\tau) \to 1$ in this case. Thus, in the limit where the ingress--egress process completely overtakes the spatial process, the stationary-state isogloss density is given by the parabola $\sigma = 2\rho (1-\rho)$, indicating complete independence of the individual spins.

\section{Model: undirected horizontal error}\label{app:generalization}

We now drop the requirement that $p'_I$ and $p'_E$ be zero, but retain the assumption that $p'_I = p'_E$, i.e.~that errors in the horizontal process are equally likely to innovate a feature as they are to lose it.

\subsection{Spin flip probability}
The spin flip rate associated with the (vertical) dynamics within a site is not affected by the generalization, and remains as before
\begin{equation}
  A(\vec x) = \frac{1 - s(\vec x)}{2} p_I + \frac{1 + s(\vec x)}{2} p_E
  = \frac{1}{2} \left[p_I + p_E + (p_E - p_I)s(\vec x)\right].
\end{equation}
The rate with which the spin at site $\vec x$ flips due to interactions with one of the neighbours changes though due to the modification of the model. We now have
\begin{equation}\label{eq:bmod}
\begin{split}
  B(\vec x) &= \frac{1}{4}\sum_{\vec x'}  \left[\frac{1+s(\vec x)}{2}\frac{1+s(\vec x')}{2} p_I' + \frac{1+s(\vec x)}{2}\frac{1-s(\vec x')}{2} (1-p_I') + \right. \\
  &\quad\quad\left.+\frac{1-s(\vec x)}{2}\frac{1+s(\vec x')}{2} (1-p_E') +\frac{1-s(\vec x)}{2}\frac{1-s(\vec x')}{2} p_E'\right]. 
\end{split}
\end{equation}
The first term only contributes if $s(\vec x)=s(\vec x')=-1$. Without copying errors, no change of $s(\vec x)$ would occur. However, with probability $p_I'$ there is an error the attempted copying of $s(\vec x')=-1$, and $s(\vec x)$ flips from $s(\vec x)=-1$ to $s(\vec x)=1$. The second term contributes only when $s(\vec x)=1$ and $s(\vec x')=-1$. This only results in a flip of $s(\vec x)$ (from $1$ to $-1$) if no error is made in the copying process, i.e. if there is no ingress event. Hence the prefactor $1-p_I'$. With probability $p_I'$ the copying is associated with an ingress-type error, and $s(\vec x)$ remains at $s=-1$, and hence no flip occurs. The third and fourth terms have a similar interpretation.

We now re-organise terms in Eq.~(\ref{eq:bmod}):
\begin{equation}
    \begin{split}
  B(\vec x) &= \frac{1}{4}\sum_{\vec x'}  \left[\frac{1+s(\vec x)}{2}\frac{1-s(\vec x')}{2}+\frac{1-s(\vec x)}{2}\frac{1+s(\vec x')}{2} +\right.\\
  &\quad\quad\left.+  p_I' \left(\frac{1+s(\vec x)}{2}\frac{1+s(\vec x')}{2}  - \frac{1+s(\vec x)}{2}\frac{1-s(\vec x')}{2}\right) +\right. \\
  &\quad\quad\left.+p_E'\left(\frac{1-s(\vec x)}{2}\frac{1-s(\vec x')}{2} -\frac{1-s(\vec x)}{2}\frac{1+s(\vec x')}{2} \right)\right]  \\  
  &= \frac{1}{4}\sum_{\vec x'}  \left[\frac{1-s(\vec x)s(\vec x')}{2}+  p_I' \frac{s(\vec x')+s(\vec x)s(\vec x')}{2}+p_E'\frac{s(\vec x)s(\vec x')-s(\vec x')}{2}\right]  \\    
    &= \frac{1}{4}\sum_{\vec x'}  \left[\frac{1-s(\vec x)s(\vec x')}{2}+  (p_I' + p_E') \frac{s(\vec x)s(\vec x')}{2}+(p_I'-p_E')\frac{s(\vec x')}{2}\right]. 
    \end{split}
\end{equation}
We therefore have
\begin{equation}
  A(\vec x) = \frac{1 - s(\vec x)}{2} p_I + \frac{1 + s(\vec x)}{2} p_E
  = \frac{1}{2} \left[p_I + p_E + (p_E - p_I)s(\vec x)\right]
\end{equation}
as in the original model, and
\begin{equation}
  B(\vec x) 
  = \frac{1}{2} \left[1 - \frac{1}{4} \sum_{\vec x'}\left\{ s(\vec x)s(\vec x')- (p_I' + p_E') s(\vec x) s(\vec x')-(p_I'-p_E')s(\vec x') \right\}\right],
\end{equation}
where it is important to remember that the summation over $\vec x'$ runs over the four nearest neighbours of $\vec x$. The total rate with which the spin at site $\vec x$ flips is therefore
\begin{equation}\label{eq:spin-flip-probability_prime}
\begin{split}
  w(\vec x) &= \frac{1-q}{2} \left[p_I + p_E + (p_E - p_I)s(\vec x)\right] + \\
 &\quad\quad + \frac{q}{2} \left[1 - \frac{1}{4} \sum_{\vec x'}\left\{ s(\vec x)s(\vec x')- (p_I' + p_E') s(\vec x) s(\vec x')-(p_I'-p_E')s(\vec x') \right\}\right].
\end{split}
\end{equation}

\subsection{Stationary-state feature frequency $\rho$}\label{subsec:rhoext}

As in the more basic model, we proceed to compute the magnetization in the stationary state, or equivalently the number of spins in state $+1$ (the fraction of sites where the feature is present). Similar to Eq.~(\ref{eq:sdiff}) we have
\begin{equation}\label{eq:sdiff_prime}
    \begin{split}
  \frac{S(\vec x, t + \Delta t) - S(\vec x,t)}{1/N} &= \langle -2w(\vec x)s(\vec x) \rangle \\
  &= (1-q) \left[p_I - p_E - (p_I + p_E)S(\vec x) \right] + q\Bigg[ -S(\vec x) + \\
  &\quad\quad + \frac{1}{4} \sum_{\vec x'}\left\{ S(\vec x') -(p_I' + p_E')  S(\vec x') -(p_I'-p_E')\langle s(\vec x)s(\vec x')\rangle\right\}\Bigg].
\end{split}
\end{equation}
It is now difficult to proceed for the case of general $p_I'$ and $p_E'$. The difficulty arises from the last term in Eq.~(\ref{eq:sdiff_prime}), which contains a correlation function. This means that the equation for the mean magnetisation is not closed. Similarly, if we proceeded to calculate the correlation function, we would meet averages of objects involving three different sites.

However, we can continue if we set $p_I'=p_E'$. The problematic term then vanishes. Assuming $p_I'=p_E'$ means that errors in the copying process from neighbours occur independent of what spin state is copied. We write $\lambda=p_I'=p_E'$ and always assume $0\leq\lambda<1/2$. The case $\lambda=0$ describes error-free horizontal transmission. For $\lambda=1/2$ all information is lost in the transmission, the received state is entirely random (feature present or absent, each with probability $1/2$), and independent of the transmitted state. The restriction $\lambda<1/2$ therefore encapsulates the assumption that some information is transmitted on average. A choice of $\lambda>1/2$ would reflect situations in which there is a tendency for bits to become actively inverted in transmission (`worse than random'). We do not study these here.

Taking the sum over all sites $\vec x$ in Eq. (\ref{eq:sdiff_prime}), one now finds
\begin{equation}
  \begin{split}
  \frac{m(t + \Delta t) - m(t)}{1/N}
    &= (1-q)(p_I - p_E) - (1-q)(p_I + p_E) \frac{1}{N} \sum_{\vec x} S(\vec x) - \\
    &\quad\quad - \frac{q}{N} \sum_{\vec x} S(\vec x) + \frac{q}{4N} (1-2\lambda) \sum_{\vec x}\sum_{\vec x'} S(\vec x').
  \end{split}
\end{equation}
From this in turn we obtain
\begin{equation}
    \begin{split}
  \frac{dm}{dt} &= (1-q)\left[p_I - p_E - (p_I + p_E)m\right]-2q\lambda m \\
  &= (1-q)(p_I-p_E) + m [-(1-q)(p_I+p_E)-2q\lambda].
  \end{split}
\end{equation}
Hence in the stationary state
\begin{equation}\label{eq:emm_prime}
  m = \frac{ (p_I - p_E)}{p_I + p_E+2\frac{q}{1-q}\lambda}.
\end{equation}
Comparing this to Eq.~(\ref{eq:emm}) we note the additional term proportional to $\lambda$ in the denominator. This term drives $m$ to zero. This makes sense, as errors in the copying process from neighbouring sites `equalize' the frequencies of the two spin states present in the system. Suppose for example there is a majority of $+1$ spins. Then an attempted copying event is more likely to be one in which a $+1$ spin is copied. Errors in this process reduce the fraction of $+1$ spins, and increase the fraction of $-1$ spins.

Using $\rho = (m+1)/2$, we find from Eq.~(\ref{eq:emm_prime}):
\begin{eqnarray}\label{eq:steady-rho_prime}
  \rho = \frac{1}{2}\frac{p_I-p_E+p_I + p_E+2\frac{q}{1-q}\lambda}{p_I + p_E+2\frac{q}{1-q}\lambda} 
  = \frac{(1-q)p_I +q\lambda}{(1-q)(p_I + p_E)+2q\lambda}.
\end{eqnarray}
This quantity is manifestly between $0$ and $1$ (numerator and denominator are both non-negative, and the denominator is always greater than or equal to the numerator). For $\lambda=0$ this reduces to Eq.~(\ref{eq:steady-rho}), the formula for the simpler model. We also notice that $\rho=1/2$ when $q=1$. This makes sense, there is then only the horizontal process, and copying errors happen with the same rate $\lambda$, independently of the state that is copied (absence or presence of the feature). As mentioned above, this drives the system to equal fractions of $-1$ and $+1$ spins.

\subsection{Correlation function and isogloss density}\label{subsec:corrfunc}

We start from [see Eq.~(\ref{eq:corr_start})]
\begin{equation}
  \frac{dS(\vec x, \vec y)}{dt} = \langle -2\left[w(\vec x) + w(\vec y)\right]s(\vec x)s(\vec y)\rangle,
\end{equation}
where now
\begin{equation}\label{eq:spin-flip-probability_prime2}
\begin{split}
  w(\vec x) &= \frac{1-q}{2} \left[p_I + p_E + (p_E - p_I)s(\vec x)\right]  + \frac{q}{2} \left[1 - \frac{1}{4} \sum_{\vec x'}\left\{ s(\vec x)s(\vec x')- 2\lambda s(\vec x) s(\vec x') \right\}\right]  \\
 &= \frac{1-q}{2} \left[p_I + p_E + (p_E - p_I)s(\vec x)\right]+ \frac{q}{2} \left[1 - \frac{1}{4} (1-2\lambda)\sum_{\vec x'} s(\vec x)s(\vec x') \right].
\end{split}
\end{equation}
From this we find
\begin{equation}
  \begin{split}
    \frac{d S(\vec x, \vec y)}{dt} &= (1-q) \left[ (p_I - p_E) [S(\vec x) + S(\vec y)] - 2(p_I + p_E)S(\vec x, \vec y) \right] + \\
    &\quad\quad + q \left[ - 2S(\vec x, \vec y) + \frac{1}{4} (1-2\lambda) \sum_{\vec x'} S(\vec x', \vec y) + \frac{1}{4} (1-2\lambda)\sum_{\vec y'} S(\vec x, \vec y') \right],
  \end{split}
 \end{equation}
where the summation over $\vec y'$ is over the four nearest neighbours of $\vec y$.

As in the analysis of the simpler model we now assume translation invariance and write $C(\vec r) = C(\vec x - \vec y) = S(\vec x, \vec y)$. Then, for $\vec r \neq \vec 0$,
\begin{equation}
  \begin{split}
    \frac{dC(\vec r)}{dt} &= (1-q) \left[ (p_I - p_E) [S(\vec x) + S(\vec y)] - 2(p_I + p_E) C(\vec r) \right] + \\
    &\quad\quad + q \left[ -2C(\vec r) + \frac{1}{4} (1-2\lambda)\sum_{\vec x'} C(\vec x' - \vec y) + \frac{1}{4}(1-2\lambda)\sum_{\vec y'} C(\vec x - \vec y') \right].
  \end{split}
\end{equation}
Adding and subtracting a term (the aim is to isolate a lattice Laplacian) we obtain
\begin{equation}
  \begin{split}
    \frac{dC(\vec r)}{dt} &= (1-q) \left[ (p_I - p_E) [S(\vec x) + S(\vec y)] - 2\left(p_I + p_E+2\frac{q}{1-q}\lambda\right) C(\vec r) \right] + \\
    &\quad\quad + q \left[ -2(1-2\lambda)C(\vec r) + \frac{1}{4} (1-2\lambda)\sum_{\vec x'} C(\vec x' - \vec y) + \frac{1}{4}(1-2\lambda)\sum_{\vec y'} C(\vec x - \vec y') \right].
  \end{split}
\end{equation}
Due to translation invariance the two summations on the RHS coincide, and we have (always restricting $\vec r \neq \vec 0$)
\begin{equation}\label{eq:pair-correlation-function_prime}
\begin{split}
  \frac{dC(\vec r)}{dt} &= (1-q) \left[ (p_I - p_E) [S(\vec x) + S(\vec y)] - 2\left(p_I + p_E+2\frac{q}{1-q}\lambda\right)C(\vec r) \right] + \\
  &\quad\quad + 2q(1-2\lambda)\Delta C(\vec r),
  \end{split}
\end{equation}
where $\Delta$ is the lattice Laplacian.

In the stationary state, $S(\vec x) + S(\vec y) = 2m$ and $\frac{dC(\vec r)}{dt} = 0$. Assuming $\vec r \neq \vec 0$, we then obtain the following from Eq.~(\ref{eq:pair-correlation-function_prime}):
\begin{equation}
  0 = (1-q) \left[ 2(p_I - p_E)m - 2\left(p_I + p_E+2\frac{q}{1-q}\lambda\right)C(\vec r) \right] + 2q(1-2\lambda) \Delta C(\vec r).
\end{equation}
In other words,
\begin{equation}\label{eq:before-alpha2}
  \begin{split}
    0 &= (1-q)\left(p_I + p_E+2\frac{q}{1-q}\lambda\right)\left[ m^2 - C(\vec r) \right] + q (1-2\lambda)\Delta C(\vec r).
  \end{split}
\end{equation}
In the special case $q = 0$ (i.e., no spatial interaction between neighbouring sites), this implies $C(\vec r) = m^2$ for all $\vec r \neq \vec 0$, as in the simpler model (but note that the stationary value of $m$ is different in the two models). 

We note that the term containing the lattice Laplacian does not contribute when either $q=0$ or $\lambda=1/2$. If $q=0$, then there is no copying process from neighbouring sites in the model from the beginning, so there is no interaction at all between the different sites. As mentioned above, if $\lambda=1/2$ no information is transferred in copying processes from neighbours, as the state to be copied is transferred faithfully with probability or inverted, each with probability $1/2$. This does not constitute any meaningful interaction, and neither does it create correlations between sites. The absence of the Laplacian reflects this.

In the following we will always exclude such pathological cases, and assume  $0 < q < 1$ and $\lambda\neq \frac{1}{2}$. Dividing Eq.~(\ref{eq:before-alpha2}) by $q(1-2\lambda)$ we obtain
\begin{equation}\label{eq:to-be-solved-2}
  \begin{split}
    0 &= \frac{q-1}{q(1-2\lambda)}\left(p_I + p_E+2\frac{q}{1-q}\lambda\right)\ \left[C(\vec r)-m^2\right] + \Delta C(\vec r) \\
    &= \alpha C(\vec r) - \alpha m^2 + \Delta C(\vec r),
  \end{split}
\end{equation}
with
\begin{equation}\label{eq:alpha_prime}
  \alpha = \frac{q-1}{q(1-2\lambda)}\left(p_I + p_E+2\frac{q}{1-q}\lambda\right).
\end{equation}
This is of the same form as Eq.~(\ref{eq:to-be-solved}), except that the expression for $\alpha$ has changed [cf.~Eq.~(\ref{eq:alpha})]. 

We can therefore use all results obtained for the simpler model, provided we use the expression for $\alpha$ in Eq.~(\ref{eq:alpha_prime}). In particular we have
\begin{equation}\label{eq:sigmarho-2}
  \sigma = 2\rho (1-\rho) \left[\alpha + \frac{\pi (1-\alpha)}{2K_{\alpha}} \right].
\end{equation}
as before  [cf.~Eq.~(\ref{eq:sigmarho})], but where $\alpha$ is given in Eq.~(\ref{eq:alpha_prime}). Defining again $\tau=-\alpha$, i.e. $K_{\alpha} = K_{-\tau}$, we have again [cf.~Eq.~(\ref{eq:parabola})]:
\begin{equation}\label{eq:parabola_prime}
  \sigma = 2 H(\tau) \rho (1-\rho),
\end{equation}
where 
\begin{equation}\label{eq:Htau2}
  H(\tau) = \frac{\pi (1 + \tau)}{2K\left(\frac{1}{1+\tau}\right)} - \tau.
\end{equation}
The height of the downward-opening parabola in $\rho$ in Eq.~(\ref{eq:parabola_prime}) is fixed by $H(\tau)$, where
\begin{equation}
\tau=-\alpha=\frac{1-q}{q(1-2\lambda)}\left(p_I + p_E+2\frac{q}{1-q}\lambda\right),
\end{equation}
i.e.,
\begin{equation}\label{eq:tempext}
\tau= \frac{(1-q)(p_I + p_E)+2q\lambda}{q(1-2\lambda)}.
\end{equation}
This expression for temperature $\tau$ clearly generalizes our earlier result (\ref{eq:tau}) for the special case $p'_I = p'_E = 0$.

We recall that the meaningful range of $\lambda$ is $0\leq \lambda<1/2$, so the temperature $\tau$ in Eq.~(\ref{eq:tempext}) is non-negative.

\section{Model: no restrictions on error rates}\label{app:full-generality}

As discussed in Section \ref{subsec:rhoext} above, we have been unable to find a closed-form solution for the stationary-state feature frequency $\rho$, isogloss density $\sigma$ and temperature $\tau$ for the fully general model, i.e.~when all of the parameters $p_I$, $p_E$, $p'_I$, $p'_E$ and $q$ are in free variation. The root of the problem lies in the fact that the equations for the first and second moments of the distribution of features do not close, when $p_I'\neq p_E'$. In such cases there is either a tendency towards introducing a feature in the horizontal process, or towards losing it, amounting effectively to selection for or against the feature. Truncation schemes would be required to obtain approximations for the stationary values for $\rho$ and $\sigma$. Interestingly very similar issues are known in the population genetic literature, in the presence of selection (\citeKorolev). We have not attempted to explore approximations for $\rho$ and $\sigma$ based on truncations of the resulting hierarchy of equations. However, the following ansatz is reasonable, and, as we show below, sufficiently accurate for our purposes.

In the model with $p'_I = p'_E$, the stationary-state feature frequency reads [see Eq.~(\ref{eq:steady-rho_prime})]
\begin{equation}\label{eq:steady-rho_prime-2}
    \rho = \frac{(1-q)p_I + q\lambda}{(1-q)(p_I + p_E) + 2q\lambda},
\end{equation}
with $\lambda = p'_I = p'_E$, so $2\lambda = p'_I + p'_E$. Since $\rho$ is by definition the frequency of up-spins, the following generalization of (\ref{eq:steady-rho_prime-2}) suggests itself:
\begin{equation}\label{eq:steady-rho-full}
    \rho = \frac{(1-q)p_I + q p'_I}{(1-q)(p_I + p_E) + q(p_I' + p_E')}.
\end{equation}
Secondly, when $p_I' = p_E'$, the stationary-state temperature reads [see Eq.~(\ref{eq:tempext})]
\begin{equation}\label{eq:tempext-2}
\tau = \frac{(1-q)(p_I + p_E) + 2q\lambda}{q(1-2\lambda)}.
\end{equation}
Substituting again $p_I' + p_E'$ for $2\lambda$ yields the following ansatz for the general case $p_I' \neq p_E'$:
\begin{equation}\label{eq:tau-full}
\tau = \frac{(1-q)(p_I + p_E) + q(p_I' + p_E')}{q(1-(p_I' + p_E'))}.
\end{equation}
Finally, from the arguments in Section \ref{subsec:corrfunc}, we hypothesize the following form for the stationary-state isogloss density for the fully general model:
\begin{equation}\label{eq:sigma-full}
    \sigma = 2H(\tau) \rho (1-\rho),
\end{equation}
where
\begin{equation}
    H(\tau) = \frac{\pi (1+\tau)}{2K\left(\frac{1}{1+\tau}\right)} - \tau
\end{equation}
and $\tau$ is taken from Eq.~(\ref{eq:tau-full}).

\begin{figure}
    \centering
    \includegraphics[width=1.0\textwidth]{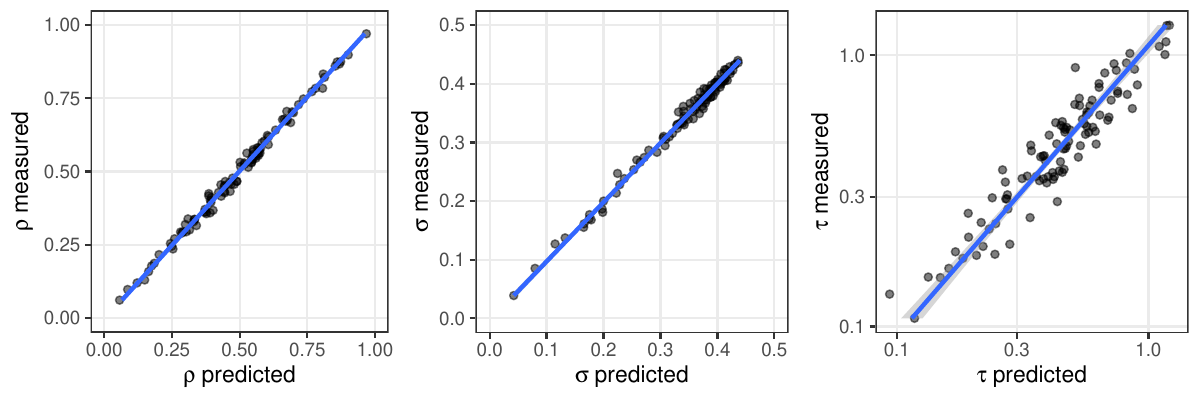}
    \caption{Correlations between quantities predicted by theory [Eqs.~(\ref{eq:steady-rho-full}), (\ref{eq:tau-full}) and (\ref{eq:sigma-full})] and measured from numerical simulations ($\rho$: feature frequency, $\sigma$: isogloss density, $\tau$: temperature). One hundred independent features on a $50\times 50$ lattice at 25,000,000 iterations.}
    \label{fig:fullmodel}
\end{figure}

To corroborate these hypotheses, we ran a simulation of the model on a lattice of $50\times 50$ sites and $100$ features, with randomly drawn values for the parameters $p_I$, $p_E$, $p_I'$, $p_E'$ and $q$ for each feature, measured the empirical values of $\rho$, $\sigma$ and $\tau$ at the end of each simulation, and correlated these against the above theoretical predictions. Each simulation was run for 25,000,000 iterations, so each feature underwent either a vertical or a horizontal event in each lattice site approximately 100 times during the simulation. The correlations are visualized in Fig.~\ref{fig:fullmodel}. The fact that all points lie close to the diagonal suggests that Equations (\ref{eq:steady-rho-full}), (\ref{eq:tempext-2}), (\ref{eq:tau-full}), even though not rigorously derived from theory, do faithfully describe these quantities at the stationary distribution. The Spearman correlation coefficients are $r_S = 0.994$ ($p < 0.001$) for $\rho$, $r_S = 0.985$ ($p < 0.001$) for $\sigma$, and $r_S = 0.931$ ($p < 0.001$) for $\tau$.

\section{Robustness of method}\label{app:robustness}

To establish that our method for estimating empirical temperatures is robust against minor variations in the data it is fed, we have repeated the analysis by applying it to disjoint subsets of the WALS (\citeWALS) dataset and by applying different spatial scales in the computation of isogloss densities, on whose values our empirical temperature estimates are based. We have also run the model on a neighbourhood graph inferred from the WALS atlas (instead of a regular lattice, as in the analytical solution), to assess to what extent the less regular real-life distribution of languages over physical space might invalidate the analytical regular-lattice idealization. Finally, in order to assess whether the irregular spatial distributions of genealogical units (language families and genera, rather than individual languages) might skew isogloss densities in unpredictable and non-universal ways, we have repeated the analysis with the restriction that nearest neighbour languages are always drawn from within the language family the focal language belongs to. The results of these extensions, which support our original idealized lattice model and the choice to consider only a small number of nearest neighbours for each language in the calculation of isogloss densities, are reported in the following subsections.

\subsection{The two hemispheres test}\label{sec:OWNW}

Our main analysis relies on a bootstrap scheme to arrive at estimates of temperature $\tau$. To make sure that $\tau$ is actually catching a signal rather than the particularities of an arbitrary sample of the world's languages, the procedure is repeated 1,000 times and the median temperatures and bootstrap distributions reported in the main paper. The question remains, however, to what extent temperature estimates from one sample of languages correlate with estimates from another such sample---a robust method should arrive at a similar ranking of features with respect to temperature $\tau$ regardless of which set of languages is considered, provided that that set is large enough. To answer this, we follow (\citeWicHol) by first partitioning the world into two subsets which are both genealogically and areally independent to a great degree: Old World (OW), consisting of longitudes between 0° E and 180° E, and New World (NW), comprising latitudes between 0° W and 180° W. The analysis is now conducted for both subsets (OW and NW) separately, and the median $\tau$ estimates from the OW bootstrap correlated against the median $\tau$ estimates from the NW bootstrap. The Spearman correlation coefficient was found to be $r_S = 0.52$ ($p < 0.01$). This should be compared against the value $r_S = 0.51$, which is what has previously been reported for a similar two hemispheres comparison based on a phylogenetic--areal technique (\citeWicHol). This reasonably high (and statistically significant) positive correlation suggests that temperature ($\tau$) is a measurable quantity whose value is to some extent independent of the specific composition of the language sample used in its calculation.

\subsection{Varying the spatial scale}

\begin{figure}
  \centering
  \includegraphics[width=1.0\textwidth]{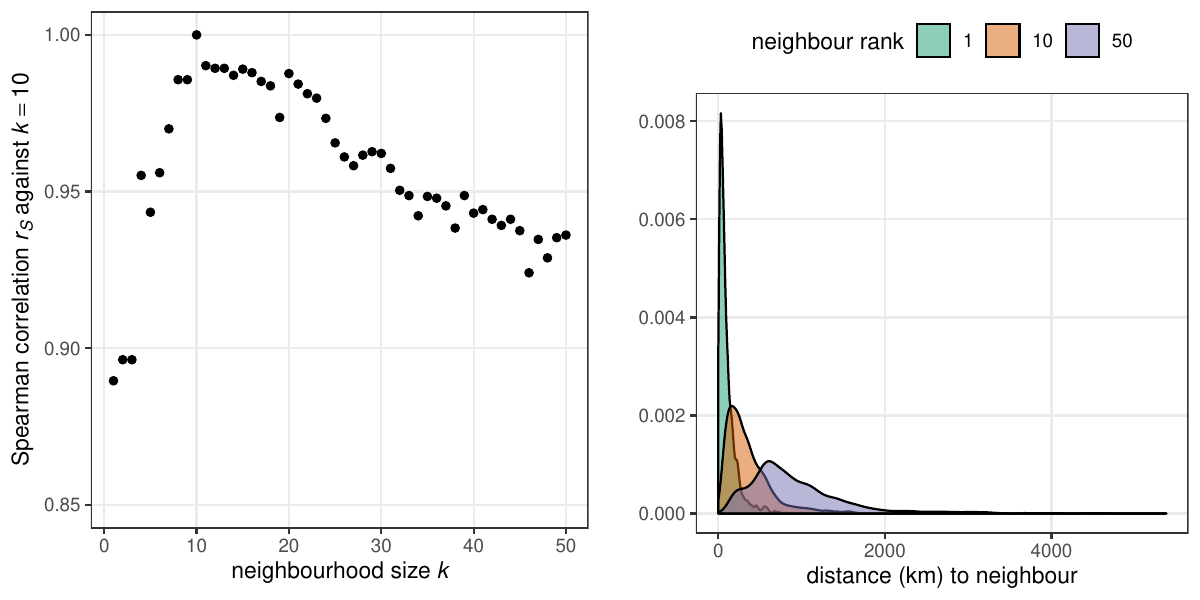}
  \caption{Left: Temperature estimates computed with varying neighbourhood sizes correlated against the temperature estimates of the main analysis (neighbourhood size $k = 10$). Right: Distribution (Gaussian kernel estimate) of distance to $k$th neighbour ($k=1, 10, 50$).}\label{fig:geoscale}
\end{figure}

In the main paper, as well as in the above two extensions, we have computed isogloss densities $\sigma$ on the basis of considering the 10 nearest geographical neighbours of each language---that is, assuming a neighbourhood size of $k = 10$. The question arises whether considering either smaller or larger neighbourhoods would noticeably affect our temperature estimates. To assess this, we repeated the analysis for neighbourhood sizes varying from $k = 1$ to $k = 50$ and correlated the temperature estimates against those found with $k = 10$ in the original analysis. The values of the Spearman correlation coefficients are plotted in Fig.~\ref{fig:geoscale} (left); all correlations are statistically significant at $p < 0.001$. The high correlations indicate that our method is robust over a range of spatial scales. Fig.~\ref{fig:geoscale} (right) shows that as neighbourhood size increases, the method becomes increasingly non-local; there is little incentive to consider larger neighbourhoods.

\subsection{Spatial adjacency graph vs.~regular lattice}

\begin{figure}
    \centering
    \includegraphics[width=1.0\textwidth]{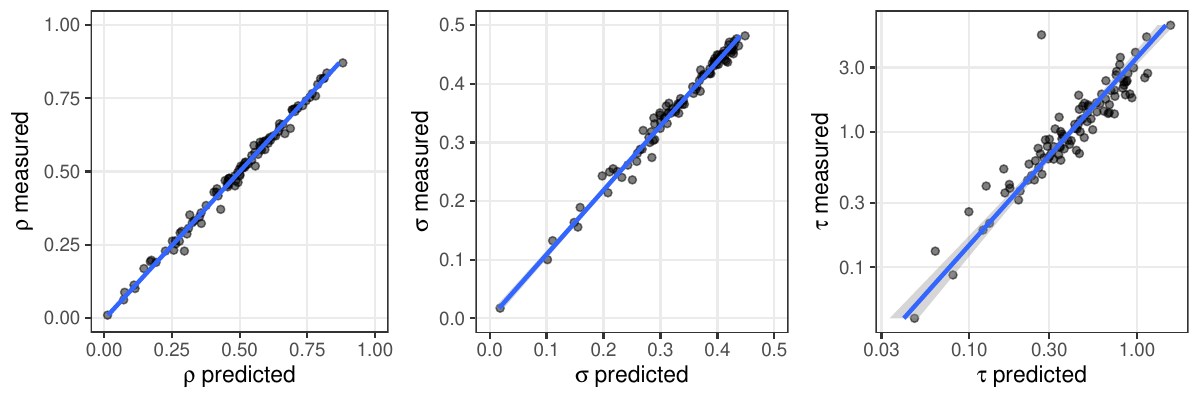}
    \caption{Correlations between quantities predicted by theory [Eqs.~(\ref{eq:steady-rho-full}), (\ref{eq:tau-full}) and (\ref{eq:sigma-full})] and measured from numerical simulations on an adjacency graph inferred from the WALS atlas ($\rho$: feature frequency, $\sigma$: isogloss density, $\tau$: temperature). One hundred independent features at 25,000,000 iterations.}
    \label{fig:adjacency-graph}
\end{figure}

The above analytical solution of the model assumes a regular isotropic infinite lattice. Empirically, languages are not distributed over physical space in this manner, so the question arises to what extent the spatial configuration of the mathematical model actually reflects reality. To assess this, we simulated the model on an adjacency graph inferred from the geographical coordinates of real languages extracted from WALS (altogether $2639$ languages, disregarding the sign languages recorded in the atlas): each language was connected to its 10 nearest geographical neighbours. To simulate a variety of features of varying temperatures, 100 independent features were modelled, drawing the parameters $p_I$, $p_E$, $p_I'$, $p_E'$ and $q$ for each feature at random. Empirical values of feature frequency $\rho$, isogloss density $\sigma$ and temperature $\tau$ were computed at the end of each simulation (after 25,000,000 iterations, so that each feature got updated in each language roughly $100$ times). The empirical quantities were then correlated against what the theory predicts on an ideal lattice. The Spearman correlation coefficients are $r_S = 0.995$ ($p < 0.001$) for $\rho$, $r_S = 0.988$ ($p < 0.001$) for $\sigma$, and $r_S = 0.915$ ($p < 0.001$) for $\tau$ (Fig.~\ref{fig:adjacency-graph}).

\subsection{Potential baseline isogloss density bias resulting from the geographical distributions of languages belonging to a genealogical unit}

It is possible to imagine a feature which is stable in the sense of having very low ingress and egress rates but being confined to a single genealogical unit (a language family or genus) whose languages happen to be scattered among languages from other genealogical units in which this feature is not present (cf.~Fig.~2C in the main paper, imagining that the colours signify both the presence of a feature as well as the genealogical unit, i.e.~presence/absence of the feature precisely carves the genealogical boundary). In such a case, our method, which does not rely on genealogical information but only looks at the empirical spatial scattering, would conceivably report an intermediate (rather than low) temperature simply because of the areal scattering of the genealogical units, whereas a genealogical method would report maximal stability.

Certain features of our method render it relatively robust to such a scenario. Our main procedure estimates features' temperatures across languages belonging to several genealogical units; moreover, in the bootstrap set-up described in the main paper, the specific geographical composition of the language sample varies from one bootstrap run to the other. However, to make the point more quantitatively, we have implemented a variation of the original analysis, in which the nearest neighbour languages are always drawn from the same language family as the focal language, as specified in the genealogical information contained in WALS. On the assumption that the proposed scattered nature of the geospatial distributions of genealogical units causes a significant problem for our method, the temperature estimates from this genetic--areal variation of the method ought not to correlate, or at least not to correlate well, with those from our main analysis. This is, however, not the case. The Spearman correlation is $0.814$, statistically significant at the $p < 0.001$ level.

\section{WALS feature levels}\label{sec:wals-details}

\setlist{noitemsep}

The following list gives the values of the WALS (\citeWALS) features considered in this study; the italicized part after each value gives its value in our binarization (\emph{present} for `feature present', \emph{absent} for `feature absent' and \emph{N/A} if the WALS value was excluded from the binarization as irrelevant). Languages attesting irrelevant values were excluded from the corresponding feature language sample for the purposes of calculating our statistics.

\subsection*{1. adpositions}

\begin{itemize}
  \item[--] WALS feature: Person Marking on Adpositions (48A)
  \item[--] Values:
  {\small
  \begin{enumerate}
    \item[1:] No adpositions (\emph{absent})
    \item[2:] No person marking (\emph{present})
    \item[3:] Pronouns only (\emph{present})
    \item[4:] Pronouns and nouns (\emph{present})
  \end{enumerate}
  }
\end{itemize}

\subsection*{2. definite marker}

\begin{itemize}
  \item[--] WALS feature: Definite Articles (37A)
  \item[--] Values:
  {\small
  \begin{enumerate}
    \item[1:] Definite word distinct from demonstrative (\emph{present})
    \item[2:] Demonstrative word used as definite article (\emph{present})
    \item[3:] Definite affix (\emph{present})
    \item[4:] No definite, but indefinite article (\emph{absent})
    \item[5:] No definite or indefinite article (\emph{absent})
  \end{enumerate}
  }
\end{itemize}

\subsection*{3. front rounded vowels}

\begin{itemize}
  \item[--] WALS feature: Front Rounded Vowels (11A)
  \item[--] Values:
  {\small
  \begin{enumerate}
    \item[1:] None (\emph{absent})
    \item[2:] High and mid (\emph{present})
    \item[3:] High only (\emph{present})
    \item[4:] Mid only (\emph{present})
  \end{enumerate}
  }
\end{itemize}

\subsection*{4. gender in independent personal pronouns}

\begin{itemize}
  \item[--] WALS feature: Gender Distinctions in Independent Personal Pronouns (44A)
  \item[--] Values:
  {\small
  \begin{enumerate}
    \item[1:] In 3rd person + 1st and/or 2nd person (\emph{present})
    \item[2:] 3rd person only, but also non-singular (\emph{present})
    \item[3:] 3rd person singular only (\emph{present})
    \item[4:] 1st or 2nd person but not 3rd (\emph{present})
    \item[5:] 3rd person non-singular only (\emph{present})
    \item[6:] No gender distinctions (\emph{absent})
  \end{enumerate}
  }
\end{itemize}

\subsection*{5. glottalized consonants}

\begin{itemize}
  \item[--] WALS feature: Glottalized Consonants (7A)
  \item[--] Values:
  {\small
  \begin{enumerate}
    \item[1:] No glottalized consonants (\emph{absent})
    \item[2:] Ejectives only (\emph{present})
    \item[3:] Implosives only (\emph{present})
    \item[4:] Glottalized resonants only (\emph{present})
    \item[5:] Ejectives and implosives (\emph{present})
    \item[6:] Ejectives and glottalized resonants (\emph{present})
    \item[7:] Implosives and glottalized resonants (\emph{present})
    \item[8:] Ejectives, implosives, and glottalized resonants (\emph{present})
  \end{enumerate}
  }
\end{itemize}

\subsection*{6. grammatical evidentials}

\begin{itemize}
  \item[--] WALS feature: Semantic Distinctions of Evidentiality (77A)
  \item[--] Values:
  {\small
  \begin{enumerate}
    \item[1:] No grammatical evidentials (\emph{absent})
    \item[2:] Indirect only (\emph{present})
    \item[3:] Direct and indirect (\emph{present})
  \end{enumerate}
  }
\end{itemize}

\subsection*{7. hand and arm identical}

\begin{itemize}
  \item[--] WALS feature: Hand and Arm (129A)
  \item[--] Values:
  {\small
  \begin{enumerate}
    \item[1:] Identical (\emph{present})
    \item[2:] Different (\emph{absent})
  \end{enumerate}
  }
\end{itemize}

\subsection*{8. hand and finger(s) identical}

\begin{itemize}
  \item[--] WALS feature: Finger and Hand (130A)
  \item[--] Values:
  {\small
  \begin{enumerate}
    \item[1:] Identical (\emph{present})
    \item[2:] Different (\emph{absent})
  \end{enumerate}
  }
\end{itemize}

\subsection*{9. indefinite marker}

\begin{itemize}
  \item[--] WALS feature: Indefinite Articles (38A)
  \item[--] Values:
  {\small
  \begin{enumerate}
    \item[1:] Indefinite word distinct from `one' (\emph{present})
    \item[2:] Indefinite word same as `one' (\emph{present})
    \item[3:] Indefinite affix (\emph{present})
    \item[4:] No indefinite, but definite article (\emph{absent})
    \item[5:] No definite or indefinite article (\emph{absent})
  \end{enumerate}
  }
\end{itemize}

\subsection*{10. inflectional morphology}

\begin{itemize}
  \item[--] WALS feature: Prefixing vs. Suffixing in Inflectional Morphology (26A)
  \item[--] Values:
  {\small
  \begin{enumerate}
    \item[1:] Little affixation (\emph{absent})
    \item[2:] Strongly suffixing (\emph{present})
    \item[3:] Weakly suffixing (\emph{present})
    \item[4:] Equal prefixing and suffixing (\emph{present})
    \item[5:] Weakly prefixing (\emph{present})
    \item[6:] Strong prefixing (\emph{present})
  \end{enumerate}
  }
\end{itemize}

\subsection*{11. inflectional optative}

\begin{itemize}
  \item[--] WALS feature: The Optative (73A)
  \item[--] Values:
  {\small
  \begin{enumerate}
    \item[1:] Inflectional optative present (\emph{present})
    \item[2:] Inflectional optative absent (\emph{absent})
  \end{enumerate}
  }
\end{itemize}

\subsection*{12. lateral consonants}

\begin{itemize}
  \item[--] WALS feature: Lateral Consonants (8A)
  \item[--] Values:
  {\small
  \begin{enumerate}
    \item[1:] No laterals (\emph{absent})
    \item[2:] /l/, no obstruent laterals (\emph{present})
    \item[3:] Laterals, but no /l/, no obstruent laterals (\emph{present})
    \item[4:] /l/ and lateral obstruent (\emph{present})
    \item[5:] No /l/, but lateral obstruents (\emph{present})
  \end{enumerate}
  }
\end{itemize}

\subsection*{13. morphological second-person imperative}

\begin{itemize}
  \item[--] WALS feature: The Morphological Imperative (70A)
  \item[--] Values:
  {\small
  \begin{enumerate}
    \item[1:] Second singular and second plural (\emph{present})
    \item[2:] Second singular (\emph{present})
    \item[3:] Second plural (\emph{present})
    \item[4:] Second person number-neutral (\emph{present})
    \item[5:] No second-person imperatives (\emph{absent})
  \end{enumerate}
  }
\end{itemize}

\subsection*{14. order of adjective and noun is AdjN}

\begin{itemize}
  \item[--] WALS feature: Order of Adjective and Noun (87A)
  \item[--] Values:
  {\small
  \begin{enumerate}
    \item[1:] Adjective-Noun (\emph{present})
    \item[2:] Noun-Adjective (\emph{absent})
    \item[3:] No dominant order (\emph{N/A})
    \item[4:] Only internally-headed relative clauses (\emph{N/A})
  \end{enumerate}
  }
\end{itemize}

\subsection*{15. order of degree word and adjective is DegAdj}

\begin{itemize}
  \item[--] WALS feature: Order of Degree Word and Adjective (91A)
  \item[--] Values:
  {\small
  \begin{enumerate}
    \item[1:] Degree word-Adjective (\emph{present})
    \item[2:] Adjective-Degree word (\emph{absent})
    \item[3:] No dominant order (\emph{N/A})
  \end{enumerate}
  }
\end{itemize}

\subsection*{16. order of genitive and noun is GenN}

\begin{itemize}
  \item[--] WALS feature: Order of Genitive and Noun (86A)
  \item[--] Values:
  {\small
  \begin{enumerate}
    \item[1:] Genitive-Noun (\emph{present})
    \item[2:] Noun-Genitive (\emph{absent})
    \item[3:] No dominant order (\emph{N/A})
  \end{enumerate}
  }
\end{itemize}

\subsection*{17. order of numeral and noun is NumN}

\begin{itemize}
  \item[--] WALS feature: Order of Numeral and Noun (89A)
  \item[--] Values:
  {\small
  \begin{enumerate}
    \item[1:] Numeral-Noun (\emph{present})
    \item[2:] Noun-Numeral (\emph{absent})
    \item[3:] No dominant order (\emph{N/A})
    \item[4:] Numeral only modifies verb (\emph{N/A})
  \end{enumerate}
  }
\end{itemize}

\subsection*{18. order of object and verb is OV}

\begin{itemize}
  \item[--] WALS feature: Order of Object and Verb (83A)
  \item[--] Values:
  {\small
  \begin{enumerate}
    \item[1:] OV (\emph{present})
    \item[2:] VO (\emph{absent})
    \item[3:] No dominant order (\emph{N/A})
  \end{enumerate}
  }
\end{itemize}

\subsection*{19. order of subject and verb is SV}

\begin{itemize}
  \item[--] WALS feature: Order of Subject and Verb (82A)
  \item[--] Values:
  {\small
  \begin{enumerate}
    \item[1:] SV (\emph{present})
    \item[2:] VS (\emph{absent})
    \item[3:] No dominant order (\emph{N/A})
  \end{enumerate}
  }
\end{itemize}

\subsection*{20. ordinal numerals}

\begin{itemize}
  \item[--] WALS feature: Ordinal Numerals (53A)
  \item[--] Values:
  {\small
  \begin{enumerate}
    \item[1:] None (\emph{absent})
    \item[2:] One, two, three (\emph{present})
    \item[3:] First, two, three (\emph{present})
    \item[4:] One-th, two-th, three-th (\emph{present})
    \item[5:] First/one-th, two-th, three-th (\emph{present})
    \item[6:] First, two-th, three-th (\emph{present})
    \item[7:] First, second, three-th (\emph{present})
    \item[8:] Various (\emph{present})
  \end{enumerate}
  }
\end{itemize}

\subsection*{21. passive construction}

\begin{itemize}
  \item[--] WALS feature: Passive Constructions (107A)
  \item[--] Values:
  {\small
  \begin{enumerate}
    \item[1:] Present (\emph{present})
    \item[2:] Absent (\emph{absent})
  \end{enumerate}
  }
\end{itemize}

\subsection*{22. plural}

\begin{itemize}
  \item[--] WALS feature: Coding of Nominal Plurality (33A)
  \item[--] Values:
  {\small
  \begin{enumerate}
    \item[1:] Plural prefix (\emph{present})
    \item[2:] Plural suffix (\emph{present})
    \item[3:] Plural stem change (\emph{present})
    \item[4:] Plural tone (\emph{present})
    \item[5:] Plural complete reduplication (\emph{present})
    \item[6:] Mixed morphological plural (\emph{present})
    \item[7:] Plural word (\emph{present})
    \item[8:] Plural clitic (\emph{present})
    \item[9:] No plural (\emph{absent})
  \end{enumerate}
  }
\end{itemize}

\subsection*{23. possessive affixes}

\begin{itemize}
  \item[--] WALS feature: Position of Pronominal Possessive Affixes (57A)
  \item[--] Values:
  {\small
  \begin{enumerate}
    \item[1:] Possessive prefixes (\emph{present})
    \item[2:] Possessive suffixes (\emph{present})
    \item[3:] Prefixes and suffixes (\emph{present})
    \item[4:] No possessive affixes (\emph{absent})
  \end{enumerate}
  }
\end{itemize}

\subsection*{24. postverbal negative morpheme}

\begin{itemize}
  \item[--] WALS feature: Postverbal Negative Morphemes (143F)
  \item[--] Values:
  {\small
  \begin{enumerate}
    \item[1:] VNeg (\emph{present})
    \item[2:] [V-Neg] (\emph{present})
    \item[3:] VNeg\&[V-Neg] (\emph{present})
    \item[4:] None (\emph{absent})
  \end{enumerate}
  }
\end{itemize}

\subsection*{25. preverbal negative morpheme}

\begin{itemize}
  \item[--] WALS feature: Preverbal Negative Morphemes (143E)
  \item[--] Values:
  {\small
  \begin{enumerate}
    \item[1:] NegV (\emph{present})
    \item[2:] [Neg-V] (\emph{present})
    \item[3:] NegV\&[Neg-V] (\emph{present})
    \item[4:] None (\emph{absent})
  \end{enumerate}
  }
\end{itemize}

\subsection*{26. productive reduplication}

\begin{itemize}
  \item[--] WALS feature: Reduplication (27A)
  \item[--] Values:
  {\small
  \begin{enumerate}
    \item[1:] Productive full and partial reduplication (\emph{present})
    \item[2:] Full reduplication only (\emph{present})
    \item[3:] No productive reduplication (\emph{absent})
  \end{enumerate}
  }
\end{itemize}

\subsection*{27. question particle}

\begin{itemize}
  \item[--] WALS feature: Position of Polar Question Particles (92A)
  \item[--] Values:
  {\small
  \begin{enumerate}
    \item[1:] Initial (\emph{present})
    \item[2:] Final (\emph{present})
    \item[3:] Second position (\emph{present})
    \item[4:] Other position (\emph{present})
    \item[5:] In either of two positions (\emph{present})
    \item[6:] No question particle (\emph{absent})
  \end{enumerate}
  }
\end{itemize}

\subsection*{28. shared encoding of nominal and locational predication}

\begin{itemize}
  \item[--] WALS feature: Nominal and Locational Predication (119A)
  \item[--] Values:
  {\small
  \begin{enumerate}
    \item[1:] Different (\emph{absent})
    \item[2:] Identical (\emph{present})
  \end{enumerate}
  }
\end{itemize}

\subsection*{29. tense-aspect inflection}

\begin{itemize}
  \item[--] WALS feature: Position of Tense-Aspect Affixes (69A)
  \item[--] Values:
  {\small
  \begin{enumerate}
    \item[1:] Tense-aspect prefixes (\emph{present})
    \item[2:] Tense-aspect suffixes (\emph{present})
    \item[3:] Tense-aspect tone (\emph{present})
    \item[4:] Mixed type (\emph{present})
    \item[5:] No tense-aspect inflection (\emph{absent})
  \end{enumerate}
  }
\end{itemize}

\subsection*{30. tone}

\begin{itemize}
  \item[--] WALS feature: Tone (13A)
  \item[--] Values:
  {\small
  \begin{enumerate}
    \item[1:] No tones (\emph{absent})
    \item[2:] Simple tone system (\emph{present})
    \item[3:] Complex tone system (\emph{present})
  \end{enumerate}
  }
\end{itemize}

\subsection*{31. uvular consonants}

\begin{itemize}
  \item[--] WALS feature: Uvular Consonants (6A)
  \item[--] Values:
  {\small
  \begin{enumerate}
    \item[1:] None (\emph{absent})
    \item[2:] Uvular stops only (\emph{present})
    \item[3:] Uvular continuants only (\emph{present})
    \item[4:] Uvular stops and continuants (\emph{present})
  \end{enumerate}
  }
\end{itemize}

\subsection*{32. velar nasal}

\begin{itemize}
  \item[--] WALS feature: The Velar Nasal (9A)
  \item[--] Values:
  {\small
  \begin{enumerate}
    \item[1:] Initial velar nasal (\emph{present})
    \item[2:] No initial velar nasal (\emph{present})
    \item[3:] No velar nasal (\emph{absent})
  \end{enumerate}
  }
\end{itemize}

\subsection*{33. verbal person marking}

\begin{itemize}
  \item[--] WALS feature: Alignment of Verbal Person Marking (100A)
  \item[--] Values:
  {\small
  \begin{enumerate}
    \item[1:] Neutral (\emph{absent})
    \item[2:] Accusative (\emph{present})
    \item[3:] Ergative (\emph{present})
    \item[4:] Active (\emph{present})
    \item[5:] Hierarchical (\emph{present})
    \item[6:] Split (\emph{present})
  \end{enumerate}
  }
\end{itemize}

\subsection*{34. voicing contrast}

\begin{itemize}
  \item[--] WALS feature: Voicing in Plosives and Fricatives (4A)
  \item[--] Values:
  {\small
  \begin{enumerate}
    \item[1:] No voicing contrast (\emph{absent})
    \item[2:] In plosives alone (\emph{present})
    \item[3:] In fricatives alone (\emph{present})
    \item[4:] In both plosives and fricatives (\emph{present})
  \end{enumerate}
  }
\end{itemize}

\subsection*{35. zero copula for predicate nominals}

\begin{itemize}
  \item[--] WALS feature: Zero Copula for Predicate Nominals (120A)
  \item[--] Values:
  {\small
  \begin{enumerate}
    \item[1:] Impossible (\emph{absent})
    \item[2:] Possible (\emph{present})
  \end{enumerate}
  }
\end{itemize}
